\documentclass[twocolumn]{aa}
\usepackage{graphicx} 
\usepackage[colorlinks=true]{hyperref} 
\hypersetup{
linkcolor=cyan,
citecolor=blue, 
urlcolor=blue
}
\usepackage{xcolor} 
\usepackage{txfonts}
\usepackage{lipsum}
\usepackage{subcaption}         

\definecolor{colorNN}{rgb}{0.8,0.2,0.4}

\definecolor{colorMC}{rgb}{1.0,0.0,0.1}

\newcommand{\Tone}{T_{\rm 1\,\rm bar}}
\newcommand{\Teff}{T_{\rm eff}}
\newcommand{\Leff}{L_{\mathrm{eff}}}
\newcommand{\Lint}{L_{\mathrm{int}}}
\newcommand{\tpds}{t_{\mathrm{pds}}}
\newcommand{\Teq}{T_{\rm eq}}
\newcommand{\Zatm}{Z_{\rm atm}}

\newcommand{\PZ}{P_{\rm Z}}
\newcommand{\TZ}{T_{\rm Z}}
\newcommand{\RN}{R_{\rm Nep.}}
\newcommand{\RU}{R_{\rm Ura.}}
\newcommand{\mmol}{m_{\rm mol}}
\newcommand{\lambdath}{\lambda_{\rm th}}
\newcommand{\mHe}{m_{\rm He}}
\newcommand{\NHe}{N_{\rm He}}

\newcommand{\mH}{m_{\rm H}}
\newcommand{\NH}{N_{\rm H}}
\newcommand{\mHzwei}{m_{\rm H2}}
\newcommand{\NHzwei}{N_{\rm H2}}
\newcommand{\NHstar}{N_{\rm H}^{\star}}
\newcommand{\Zwaterzwei}{Z_{\rm H2O}^{(2)}}
\newcommand{\Zwater}{Z_{\rm H2O}}
\newcommand{\Zmeth}{Z_{\rm CH4}}
\newcommand{\xwaterzwei}{x_{\rm H2O}^{(\rm sep)}}
\newcommand{\Mcore}{M_{\rm core}}
\newcommand{\Req}{R_{\rm eq}}
\newcommand{\ME}{M_{\rm E}}

\begin{document}

\title{Thermal evolution of Uranus and Neptune:\\ Hydrogen–water phase separation and the role of methane}
\author{MCA, NT, nn}
    \author{Marina Cano Amoros \inst{1,2}, 
     Nicola Tosi\inst{1}, 
     Nadine Nettelmann\inst{2}}
\institute{
    Institute of Space Research, German Aerospace Center (DLR),
        Rutherfordstr. 2, 12489, Berlin,  Germany
    \and
    Institut für Physik, Universität Rostock, Albert-Einstein Str. 23-24, 18051, Rostock, Germany 
}

\abstract
{Adiabatic evolution models of Uranus and Neptune with constant compositional stratification miss their observed luminosities: Uranus is fainter than commonly predicted, while improved equations of state (EoS) have recently led to a brightness problem for Neptune. In adiabatic models of Jupiter and Saturn, helium rain is thought to enhance luminosity. In the ice giants, hydrogen--water phase separation (H$_2$--H$_2$O demixing) with water rain has recently been predicted.}
{We aim to understand how H$_2$--H$_2$O demixing and the EoS of water influence the luminosity of Uranus and Neptune.}
{We use state-of-the-art EoS of H, He, H$_2$O, and the H$_2$--H$_2$O phase diagram to build adiabatic three-layer models of the present-day structure that match the observed gravity field and the phase diagram. We compute evolution tracks with and without demixing, and for Neptune, with and without latent heat and methane, for which we provide an updated large-scale EoS including entropy.}
{Without demixing, adiabatic Neptune with the revised AQUA EoS, without methane or latent heat, cools to its observed luminosity within 4.5 Gyr, in good agreement with the age of the Solar system. Applying the same assumptions to Uranus would require a Bond albedo of 0.43 to explain its luminosity. With demixing, the radii expand while the luminosity decreases. For Uranus, demixing leads to an equilibrium evolution with the incident flux that naturally explains its faintness. Demixing also makes Neptune fainter. Methane reduces the radius expansion but not the faintness problem. Favored Neptune models include demixing and methane at an initial C/O ratio of $\sim$1 due to CO ice accretion, with cooling times of $\lesssim 4$~Gyr.}
{Neptune requires an additional heat source unless its structure is primordial and the outer envelope is ice-poor, while hydrogen-water phase separation is an option to explain the structure and faintness of Uranus.}

\authorrunning{Cano Amoros et al.}
\titlerunning{Evolution of Uranus and Neptune}

\maketitle

\section{Introduction}

Uranus and Neptune are remarkably similar in mass, radius, and bulk composition, but their intrinsic heat fluxes differ by nearly an order of magnitude. Based on Voyager radiometry, Uranus was found to emit an intrinsic flux of $42\pm47$~erg~s$^{-1}$~cm$^{-2}$, consistent with zero within the observational uncertainty \citep{PEARL1990}, whereas Neptune emits $433\pm46$~erg~s$^{-1}$~cm$^{-2}$ \citep{pearlconrath1991}. 
More recent analyses of Uranus' radiative energy budget by \citet{wang2025} and \citet{irwin2025} have provided evidence for a non-zero internal heat flux of Uranus, although still substantially lower than that of Neptune. This revised estimate therefore does not remove the large difference in intrinsic heat flux between the two planets. Instead, it reinforces the importance of understanding why otherwise similar planets appear to undergo different thermal evolutions. As emphasized by \citet{stevenson2026}, the puzzle is not necessarily the faintness of Uranus but the difference between the two planets.

Early thermal evolution calculations that treated Uranus and Neptune as adiabatic with a layered interior structure could more or less reproduce Neptune's luminosity but found Uranus  too luminous compared to observations \citep{Hub1977, HMacF80, hubbard1995, Fortney_2011, nett2013}.  Neptune seemed to be the "standard" adiabatically-cooling planet until \citet{scheibe19} showed that using updated equations of state (EoS) for hydrogen, helium, and water, neither Uranus nor Neptune can be explained with a purely adiabatic interior because Uranus would still cool too slowly while Neptune too fast. 
\citet{scheibe19} identified the updated H/He-EoS to be responsible for the faster cooling, while the water EoS of \cite{Mazevet2019} that was later incorporated into original AQUA \citep{aqua2020} would yield similar results as the H$_2$O-REOS \citep{Nettelmann08} water EoS.

Alternatives to the planets cooling adiabatically were investigated,
especially tackling the issue of Uranus' faintness. \citet{podolak1991} and \citet{hubbard1995} suggested convection may be inhibited in the interiors due to a gradient in composition, or that the initial thermal energy budget of Uranus was low.  \citet{nettelmann16} assumed inhibition of convection in Uranus as a result of a permanent compositional gradient between an H/He-rich outer envelope and the ice-rock-rich lower envelope. The thermal boundary layer (TBL) that was assumed to grow there over time traps heat inside whilst letting the outer envelope cool down faster. Such a TBL can explain the faintness of Uranus but not the brightness of Neptune. \citet{scheibe21} followed up on the assumption of a permanent compositional gradient of given thickness and thermal conductivity  and studied a TBL that would evolve self-consistently with the interior heat flux. They found this  setup can lead to either a deficit or an excess luminosity relative to the adiabatic case over time so that slightly different parameter values could explain both planets' luminosities. 

\citet{vazan20} showed that a deep, primordial compositional gradient can indeed persist for several billions of years and explain the low luminosity of Uranus if, in addition, its initial energy budget was not much more than its present one.  

\citet{tejada-arevalo2025} modeled the evolution of initial compositional gradients with convective mixing similar  to \citet{vazan20}, but in addition, accounted for the possible presence of methane and ammonia. Measurements indicate a high CH$_4$/H$_2$ ratio of 50 to 90 $\times$ solar in the atmospheres of both planets \citep{Atreya20}.
\citet{tejada-arevalo2025} confirmed that different initial entropies matter for the long-term stability of primordial gradients. To explain the current brightness of Neptune, they suggested a higher initial entropy. This way, the outer compositional gradient would become convectively unstable, leading to mixing and adiabatic cooling at high luminosity and to a higher outer envelope metallicity at the present day for Neptune. These results highlight the inter-dependence of thermal and compositional evolution.  

Hydrogen–water phase separation in a water-rich planet is a mechanism that counteracts the homogenizing effect of convective mixing. Moreover, \citet{Bailey_2021} suggested that the gravitational energy release from water rain-out associated with H$_2$--H$_2$O  demixing  could deliver the excess luminosity of present-day Neptune, while Uranus's interior may be already fully demixed and therefore no longer release gravitational energy from this process. In contrast,
\citet{Gupta25} suggest that hydrogen–water phase separation occurs in both Uranus and Neptune at the present-day.

\citet{canoamoros2024} investigated the influence of H$_2$--H$_2$O demixing on the water abundance and interior structure of Uranus and Neptune by applying experimental and numerical predictions on the hydrogen-water phase diagrams. They showed that demixing would proceed similarly in both planets as a function of the atmospheric temperature, and since both have similar observed temperatures at 1-bar of $\sim$76 K and $\sim$72 K, the atmospheric water depletion would be similar today as well or even more pronounced in Neptune  due to its slightly cooler atmosphere. The different heat flows of the two planets remained an open issue.

\citet{HowardHelledAA24} subsequently coupled  H$_2$--H$_2$O demixing with planetary thermal evolution. They accounted for a large uncertainty in both the H$_2$--H$_2$O phase diagram and the water EoS through an offset in temperature of the phase diagram, which they constructed from the DFT-MD data of \citet{berg24}. They showed that a strong water depletion for high assumed demixing temperatures can lead to a radius inflation.  
Under their assumptions, either Neptune would cool too fast or both planets end-up too big. This suggests that a shift in temperature alone does not cover the full uncertainty in the phase diagram, in the the water EoS, in the composition, or that further processes are important.

Recently, \citet{canoamoros2024} (hereafter Paper~1) constructed a H$_2$--H$_2$O phase diagram also based on the DFT-MD data of \citet{berg24} but further constrained by experimental data. We label it \emph{Berg24}. It yields a steep water abundance difference that grows as the planet cools and is located rather far out at $\gtrsim$0.9 $M_p$. The amount of water that demixes and rains down depends on the thermal profile and thus also on the water EoS. In Paper 1, we applied the AQUA EoS \citep{aqua2020}. Afterward, \citet{canoamoros26} and independently \citet{tejada-arevalo2025} corrected for an error in the entropy of the AQUA EoS. We label the  revised version by \citet{canoamoros26} \emph{Revised AQUA}.

In this work, we revisit the thermal evolution of Uranus and Neptune. Since it has been demonstrated that the influences of the EoS of H, He, and water as well as the process of H$_2$--H$_2$O demixing on the cooling and radius evolution can be important and that methane is an abundant constituent of their atmospheres, we include here all these features. We do that based on state-of-art predictions for the material properties and observational constraints.
We apply the \emph{Berg24} phase diagram, the Revised AQUA EoS, and a wide-range methane EoS that includes entropy.
Our aim is to reassess whether additional physical processes need to be considered to explain the observed luminosities of Uranus and Neptune. For that purpose, we consider different evolution tracks with and without phase separation, with and without methane, with and without the latent heat from a phase transition to superionic water, and an uncertainty in albedo. 

The paper is organised as follows. Section \ref{sec_methods} describes our method of structure and evolution modeling. including how we map structure onto time  and the equations of state used. In Section \ref{sec:etracks} we design different tracks (T1---T6) to follow the  evolution of Uranus and Neptune. 
Section \ref{sec:results-uranus} presents the results for the evolution of Uranus and Section \ref{sec:results-neptune} of Neptune. This  includes the effect of using the original versus the revised AQUA EoS and for Neptune,  the effect of the latent heat released at the fluid-superionic transition of water, and the effect of adding methane. In Section \ref{sec_disc} we discuss the results with respect to a possible primordial structure of Neptune (\ref{sec:disc_adN}) and the Bond albedo of Uranus (\ref{sec:disc_albU}), and outline directions for future work. Section \ref{sec_conc} summarizes our conclusions. 
In a longer Appendix, we derive  the equations for the inclusion of methane (Appendix \ref{apx:meth_eqs}), describe the methane EoS with entropy (Appendix \ref{apx:methane}), which is an updated compilation of that previously applied to the icy Uranus model in \cite{Bethkenhagen17}. Appendix \ref{apx:lumi} contains additional Figures and Appendix \ref{apx:gravimoms} a numerical accuracy assessment.

\section{Methods}
\label{sec_methods}

\begin{table*}
\centering
\caption{Observational constraints used in this work for Uranus and Neptune's interior structure and evolution modelling.}
\label{tab:obsconstraints}
\begin{tabular}{lllll}
\hline\hline
 & Uranus &  & Neptune &  \\
\hline
Mass $M$ [$10^{24}\,\mathrm{kg}$] & 86.8099 & \cite{Jacobson_2025} & 102.409 & \cite{jacobson2009}\\
Equatorial radius $R_{\mathrm{eq}}$ [km] & 25 559 $\pm$4 & \cite{lindal1987} & 24 766 $\pm$ 15 & \cite{lindal1992} \\
Pressure $P$ at $R_{\mathrm{eq}}$ [bar] & 1 & \cite{lindal1987} & 1 & \cite{lindal1992} \\
$\Tone$ [K] & 76 $\pm$ 2 & \cite{lindal1987} & 72 $\pm$ 2 & \cite{lindal1992} \\
Effective temperature $T_{\mathrm{eff}}$ [K] & 59.1 $\pm$ 0.3 & \cite{PEARL1990}  & 59.3 $\pm$ 0.8 & \cite{pearlconrath1991} \\
Bond Albedo $A$ & 0.3 $\pm$ 0.049 & \cite{PEARL1990} & 0.29 $\pm$ 0.067 & \cite{pearlconrath1991}\\
Equilibrium temperature $T_{\mathrm{eq}}$ [K] & 58.2 $\pm$ 1.0 & \cite{PEARL1990} & 46.6 $\pm$ 1.1 & \cite{pearlconrath1991} \\
Rotation period $P_{\mathrm{rot}}$ [s] & 62 060 $\pm$ 40 & \cite{guillot2015} & 58 000 $\pm$ 200& \cite{guillot2015}\\
Gravitational moment $J_2$ [$\cdot 10^{6}$] & 3509.3 $\pm$ 0.41 & \cite{french2024} & 3529.4 $\pm$ 4.5 & \cite{helled2010} \\
Gravitational moment $J_4$ [$\cdot 10^{6}$] & –35.5 $\pm$  0.47 & \cite{french2024}  & –35.8 $\pm$ 2.9 & \cite{helled2010} \\
\hline
\end{tabular}
\end{table*}

\subsection{Evolution modeling}
\label{sec:meth_evol}

We compute the thermal evolution of Uranus and Neptune using a newly-developed, python-based code, 
which is now integrated into the interior structure and evolution code TATOOINE \citep{Baumeister_2020,MacKenzie_2023,Baumeister_2023}. 
The new code follows the methodology of MOGROP \citep{nett2012} and was benchmarked against it. 

First, a series of piecewise adiabatic interior structure models at different 1 bar temperatures $\Tone$ is calculated. For weakly irradiated planets, this physical temperature can be related to the effective temperature \citep{Guillo95},
\begin{equation}
    T_{\mathrm{1 bar}} = K\: T_{\mathrm{eff}}^{1.244} g^{-0.167},
\label{eq:T1bar}
\end{equation}
where $g$ is the planet's gravitational acceleration at the surface and $K$ is a fit parameter that is calibrated to yield the observed $T_{\mathrm{eff}}$. As in \cite{scheibe19}, we use the constant values for Uranus and Neptune $K_{\rm Ura}$ = 1.481 and $K_{\rm Nep}$ = 1.451, respectively, with $g$ in cm s$^{-2}$.  

The temperature $\Teff$ is equivalent to the observable luminosity  from the planet at infrared wavelengths, $\Leff$, which is to be distinguished from the total luminosity $L$ over all wavelengths. While $L$ contains contributions from the reflected light from the stellar irradiation, $\sim$$A_B\,T_{\mathrm{eq}\, 0}^4$, where $A_B$ is Bond Albedo and $T_{\mathrm{eq}\,0}$ is zero-Albedo equilibrium temperature, and also from absorbed and re-emitted stellar irradiation $L_{\rm eq}$$\sim$$(1-A_B)\,T_{\mathrm{eq}\, 0}^4$, $\Leff$ contains only the latter. 
The spectral energy distribution in $\Leff$ follows the spectrum of a black-body. For planets with $\Teff$ of the order of 100~K, it peaks at IR-wavelengths. The observable $\Leff$ yields $\Teff$ via Stefan-Boltzmann's law. That law can be generalized  for different luminosity contributions, $L_i=4\pi\, R_p^2\, \sigma_{B}\,T_i^4$, where  $\sigma_B$ is the Stefan-Boltzmann constant.

The energy balance equation to be solved reads
\begin{equation}\label{eq:Leff}
\Leff = L_{\rm eq} + \Lint,
\end{equation}
with  the equilibrium temperature $\Teq^4=(1-A_B)\,T_{\mathrm{eq}\,0}^4 = (1-A_B) (R_{\ast}^2/2a)$,
where $R_{\ast}$ is the stellar radius and $a$ is orbital distance.
$\Lint$ is the heat loss from the interior that itself can contain various contributions. We include secular cooling and the radiogenic heat loss from a rocky core, $L_{\rm radio}$,  as in \citet{Nettelmann11}. Thus,
$ L_{\rm int} =  L_{\mathrm{sec}} + L_{\rm radio}\:$. 

The secular cooling accounts for heat loss from cooling and the associated contraction of the planet. Both effects lead to a change in the gravitational energy, $dE_\mathrm{grav}$, and in the internal energy, $dE_{\mathrm{\,i}}$, of the planet. 
Secular cooling can therefore be written as $L_{\mathrm{sec}}= -(dE_{\mathrm{grav}} + dE_{\mathrm{\,i}})/dt$.
Mass re-distribution is relevant as a result of water rain-out. It influences both the gravitational and the internal energy distribution. These contributions can be merged into the heat loss $\delta q/dt= T\,ds/dt$, and therefore, the energetic effect of rain-out is implicitly included in the local changes of specific entropy $\mathrm{d}s$ over time:
\begin{equation}\label{eq:Lsec}
  L_{\mathrm{sec}}(t)  \;=\; 
  - \frac{1}{\mathrm{d}t} \int_{M_{\mathrm{core}}}^{M_{\mathrm{p}}}\!\!\!dm\:\Big( T(m, t)\, \mathrm{d}s(m) \, 
    + \, M_\mathrm{core}\,c_{v}\, \mathrm{d}T_{\mathrm{core}}(t)\:\Big)
\end{equation}
with specific heat capacity $c_{v}$, for which we use 1~J/(g K).

We pre-compute a set of 55 non-rotating interior structure profiles at $\Tone$ between 500 and 65 K. Each model provides radial profiles of mass ($m$), pressure ($P$), temperature ($T$), density ($\rho$), radius ($r$), gravity, and entropy ($s$) extending from the 1-bar level to the planetary center.  Because the interior models are computed at discrete $\Tone$ values, interpolation is used to obtain additional intermediate states to ensure a sufficient temporal resolution during the evolution.

Given the series of interior profiles ordered by $\Tone$, their local differences in entropy $ds(m)$ at the mass coordinate $m$ can be computed, and equation (\ref{eq:Leff}) solved for the time interval $dt$ between consecutive profiles. Thermal evolution is the integration over many time intervals. We let the evolution start at $t=0$ and integrate forward in time.

\subsection{Interior structure modelling}
\label{sec:meth_struc}

We obtain interior structure profiles for Uranus and Neptune using the 1D interior structure and evolution code TATOOINE. The models are constrained by  the observational quantities summarized in Table~\ref{tab:obsconstraints}.

As in Paper~1, we adopt a traditional three-layer structure with two adiabatic envelopes overlying an isothermal core \citep{HMacF80,nett2013}, noting that multi-layered approaches that account for the formation history \citep{Malamud24} or demixing deep inside \citep{Militzer_2025} may be more appropriate representations of the ice giants' deep interiors. TATOOINE solves the equations of mass conservation and hydrostatic equilibrium of non-rotating planets together with suitable equations of state \citep[e.g.][]{Baumeister_2020}.

\subsubsection{Equations of state and structure parameters}

The outer and inner envelopes are composed of a mixture of hydrogen, helium, water, and, in some cases, methane. For H and He we use the CD21 EoS of \cite{cd21}. For water we use Revised AQUA \citep{canoamoros26}. For methane we use the EoS described in Appendix \ref{apx:methane}. We calculate the EoS of mixtures via the additive volume rule. 
For the rock core we use the $P$--$\rho$ relation of \citet{hubbard1989} for a mixture of silicates, magnesium oxide, and iron compounds, which is an improved fit to that by \citet{HMacF80}. The EoS of the individual rock components are based on interpolations between theories that apply in different areas of the $P$--$T$ phase space and experimental data at very high pressures up to $\sim$25 Mbar, which were achieved by shock compression experiments using nuclear explosives \citep{Altshuler1965,Walzer1982}.

The specific entropy of the mixture with mass abundances $X$ (H), $Y$ (He), and $Z$ (heavy elements) is obtained by extending the ideal mixing prescription of \citet{chabrier19} (their Eqs.~9--11) for the binary H--He system to this multi-component system. The adiabat is then constructed by interpolation along a path of constant entropy $S(T,P,\{X_i\})$, starting from a specified temperature at 1 bar. The transition pressure and temperature between the outer and inner envelope are denoted $\PZ$ and $\TZ$, respectively. The two envelopes have different compositions with bulk heavy element mass fractions $\Zatm(=Z_1$) and $Z_2$, respectively. The inner envelope follows another adiabat defined by $S(T_Z,P_Z,\{X_{i,2}\})$ and the composition there $\{X_{i,2}\})$. As in Paper~1, for $\Tone < 100$ K and $P<10$ bar we use the simple ideal gas EoS. 

Below we describe how we determine the composition of the planet in terms of the four parameters $Z_1$, $Z_2$, $P_Z$, and $\Mcore$.

\subsubsection{Outer envelope metallicity and transition pressure}
\label{sectionZatm}

Current interior structure models of the ice giants indicate a H/He-dominated outer envelope overlying a deeper interior that is enriched in heavy elements (Helled et al. 2020).
\cite{Bailey_2021} proposed hydrogen-water immiscibility as a physical explanation for the metal-poor outer envelopes. 

In Paper~1 we constructed the hydrogen-water phase diagram labeled \emph{Berg24}. It employs experimental data of \citet{Bali13} and \citet{Vlasov23} and constraints from the DFT-MD simulations of \citet{berg24}. We use it to constrain the metallicity \(Z_1\) of the outer envelopes, as well as the transition pressure \(P_Z\) between the two envelopes. The way we do it is as follows.

To determine the particle fraction of water molecules $\xwaterzwei$ in the two-component hydrogen-water system we use  the iterative procedure between the adiabat  and the phase diagram described in Paper~1. 
The conversion between water particle and mass fraction was made as if there were no other constituents i.e.~by setting $\Zwater=\Zwaterzwei$ and using Eq. \eqref{eq.xZH2O} except if there is methane, see Section \ref{sec:etracks}. The assumed composition affects the equilibrium abundance $\xwaterzwei$, and hence $\Zwater$ through the composition-dependence of the temperature along the adiabat.
The helium mass fraction $Y$ is calculated as $Y = Y^{(2)}\,(1-Z)$ where $Y^{(2)}$  is the helium mass fraction in a two-component H-He system. We set it to a protosolar value of 0.275.

Once $\Zwater$ has been determined for a given $\Tone$ value, we take it as the water abundance in the outer envelope (except for track T5). Thereby we implicitly assume that convective overturn above the rain-out region is fast compared to the rain-out timescale, so that the entire outer envelope is homogenized to the composition at the top of the region of phase separation. Finally, $Z_1=\Zatm=\Zwater+\Zmeth$ is the total outer envelope metallicity.  

The demixing procedure also yields the transition pressure $P_Z$ between the outer and the inner envelope. We find that $P_Z$ varies only little with $\Tone$ at a value around 11 GPa. Therefore, we keep $P_Z$ constant at 11 GPa for structures where demixing occurs. For structures without demixing, the transition between inner and outer envelope is determined by the requirement of mass conservation of each layer, and $P_Z$ does not play a role.

\subsubsection{Present-day structure}
\label{section_presentday}

\begin{table*}
\centering
\caption{Present-day interior structure models for Uranus and Neptune.}
\label{tab:pds}
\begin{tabular}{llllllllll}
\hline\hline
\noalign{\vskip 3pt}
 &$\Tone$ & $\Zatm$ & $Z_{2}$  & $\PZ$ & $M_{\mathrm{core}}$ & $M_{\rm water}$ & $R_{\mathrm{eq}}$  & $J_2$ &$J_4$  \\
  &[K] &  &   &  [GPa] & [$M_{\mathrm{E}}$] & [$M_{\mathrm{p}}$ , $M_{\mathrm{E}}$] & [km] & [$\cdot 10^{6}$]  & [$\cdot 10^{6}$]\\
\noalign{\vskip 1pt}
\hline
\noalign{\vskip 3pt}
Uranus & 76 & 0.141 & 0.854 & 11 & 1.39 & 0.72 , 10.48 & 25 561 & 3509.5 & -35.9\\
\noalign{\vskip 3pt}
Neptune & 72 & 0.100 & 0.905& 11 & 1.13 & 0.81 , 13.84 & 24 769 & 3527.4 & -33.2  \\
\noalign{\vskip 3pt}
Neptune w/ CH4 & 72 & 0.179 & 0.922 & 11 & 1.46 & 0.72 , 12.37 & 24 756  & 3529.0 & -33.9  \\
\noalign{\vskip 1pt}

\hline
\end{tabular}
\end{table*}

We define \emph{present-day structure} models of Uranus and Neptune as those that satisfy the observational constraints for $M_p$, $\Req$, $\Tone$, $P_{\rm rot}$, and $J_2$ and $J_4$ given in \autoref{tab:obsconstraints}. 

To obtain the present-day structure, we use $Z_1$ and $P_Z$ of Section  \ref{sectionZatm} and adjust the two free parameters $Z_{2}$ and core mass to reproduce the observed values of the gravitational harmonics $J_2$ and $J_4$. 
To calculate $J_2$, $J_4$, and the equatorial radius $\Req$ of the rotating planets we hand the 1D structure models from TATOOINE over to the MOGROP code. We iterate between the mean radius assumed in the 1D structure models and the composition ($Z_2$, $\Mcore$) on the one hand, which is the TATOOINE part, and the values of $\Req$, $J_2$, and $J_4$ according to the theory of figures of \citet{zharkov78} up to the fourth order on the other hand, which is the MOGROP-part. 

Using the TATOOINE code for computing the structure models is a choice made for our convenience. However, we must note there are small inconsistencies at the current level of automation of the interaction between the two codes, TATOOINE and MOGROP. Those are addressed in Appendix \ref{apx:gravimoms}.

The parameter values in \autoref{tab:pds} list our results for the present-day interior structures of Uranus and Neptune. 
The values of $Z_{2}$ and $\Mcore$ differ slightly from those of equivalent models in Paper~1 because the entropy in Revised AQUA modifies the $P$--$T$ profile in the deep envelope, while the updated gravitational moments provide revised structural constraints.

The different evolutionary tracks T1--T6 we explore differ in the initial conditions, pathways, and time at which the present-day structure is reached.

\subsection{Structures in time}

The starting point in time for our evolution runs is $t=0$, with the corresponding planetary structure defined as the \emph{initial} model. This is computed for an initial temperature of $\Tone=500\:$K. This is high enough for $d\Tone/dt$ and $d\Teff/dt$ to fall quickly, i.e., further increasing $\Tone$ would not lead to a further evolution in time,
as can be seen in Figure \ref{evol-uranus-neptune} panels c,d.

The time when the present-day structure model occurs is labeld $\tpds$. In general, $\tpds$ deviates from the age of the solar system, $\tau_{\odot} = 4.56$~Gyr. If they agree, the model explains the current observed luminosity.
The time when H$_2$--H$_2$O phase separation begins is labeled $t_{\mathrm{sep,i}}$  and when it ends, $t_{\mathrm{sep,f}}$.

\(M_{\rm core}\) is kept constant over time. When demixing occurs, we also keep \(P_Z\) constant while \(Z_{1}\) and \(Z_2\) may change as a result of water rainout. Starting with the present-day structure models, we compute $Z_{2}$ at progressively higher $\Tone$ values by the requirement of water mass conservation as described in Paper~1. 
In the deep envelope, we neglect the change in methane mass fraction  with the change of the water mass fraction there.

Some evolutionary tracks can be extended beyond the time when the present-day structure is reached. This can go along with values of $\Tone$ and $\Teff$ lower than those observed today. As a consequence, $\Teff$ can decrease down to $\Teq$, after which the evolution of the atmosphere may proceed in close radiative equilibrium with the incident flux. Unfortunately, neither our codes MOGROP nor TATOOINE can model an equilibrium evolution. An equilibrium evolution ($\Teff \gtrsim \Teq$) can be modeled under the assumption of a TBL, where heat loss from the deep interior provides sufficient energy to delay the cooling of the outer envelope or even to cause it to warm-up so that $\Tone$ can increase with time as $\Teff$ follows the increasing insolation $\Teq$ \citep{scheibe21}. Here, we simply stop the evolution right before that happens. 

\subsection{Evolution tracks}
\label{sec:etracks}

We consider three different evolutionary tracks for Uranus, and six for Neptune. Tracks 1 and 6 do not include H$_2$-H$_2$O demixing; tracks 2---5 do. Tracks 1---3 assume envelopes that consist of H, He, and water. Tracks 4---6 include methane.
\begin{enumerate}
\item [T1] Track 1 assumes that no demixing occurs. T1 represents a scenario in which the compositional stratification ($Z_1$, $Z_2$, $\Mcore$) remains primordial. 

\item [T2]
Track 2 represents an initially quite homogeneous planet with an early onset of demixing. The initial $\Zatm$ is set to 0.7.
To construct this track, we begin from the present-day structure $\{\Zatm,P_Z,Z_2,\Mcore\}$ and determine earlier compositions by progressively increasing $\Tone$ while reversing the effects of demixing until the outer-envelope abundance reaches $\Zatm=0.7$, after which the abundances are kept constant.  

\item[T3]
Track 3 is similar to T2 but with a lower initial $\Zatm=0.31$, leading to a later onset of demixing. 

\item[T4]
Track 4 assumes a constant methane mass fraction $\Zmeth$ throughout the envelopes and over time. Its value is found under the two imposed conditions: first, the water abundance on the outer envelope of the present-day model is consistent with the phase diagram and second, the gravitational harmonics are matched (although see Appendix \ref{apx:gravimoms}). 
A further, and as it turns out, very strong constraint is $\Zmeth=\Zwater$ in the outer envelope at $t=0$. 
The motivation is that here, we assume that carbon and oxygen were originally accreted in form of CO ice. Since the  molecular weights of CH$_4$ and H$_2$O are quite similar, so would be their initial mass fractions.

\item[T5]
Track 5 assumes constant $\rm CH_4/H_2$  and He/H particle fractions in the outer envelope over time . This leads to a change of their mass fractions when $x_{\rm H2O}$ changes. At all times, the water mass fraction $\Zwater$ in the outer envelope is adjusted to the $x_{\rm H2O}$ according to the phase diagram and the presence of methane and helium, see Appendix \ref{apx:meth_eqs}. As for T4, the present-day structure assumes a constant methane mass fraction throughout both envelopes.

\item[T6]
Track 6 is with methane but without demixing. The present-day structure of T4 is assumed and thermally evolved.   
\end{enumerate}

\section{Results for Uranus}
\label{sec:results-uranus}

\begin{table*}
\centering
\caption{Results for evolution tracks for Uranus at different times. }
\label{tab:uranus}
\begin{tabular}{c c c c c c c c c c c}
\hline
  & $Z_1(0)$ & $Z_2(0)$ & $Z_1(t_{\rm pds})$   & $t_{\rm pds}$
  & $t_{\mathrm{sep},\,i}$ & $\Tone(t_{\mathrm{sep},\,i})$ & $t_{\rm eq}$ & $\Tone(t_{\rm eq})$ & $R_p(t_{\rm eq})$\\
  & & & & (Gyr) & (Gyr) & (K) & (Gyr) & (K) & ($\RU$)\\
\hline
$T_1$ & 0.141  & 0.854 &  0.141 &  8.50  & --    & --   & 8.5  & 76 & 1.000 \\
$T_2$ & 0.700  & 0.806 &  0.141 &  1.27  & 0.11  & 271  & 2.71 & 72 & 1.005\\
$T_3$ & 0.314  & 0.839 &  0.141 &  2.68  & 1.72  & 91   & 3.32 & 72 & 1.003 \\
\hline
\end{tabular}
\end{table*}

\begin{figure*}[h]
\centering
\includegraphics[width=0.80\textwidth]{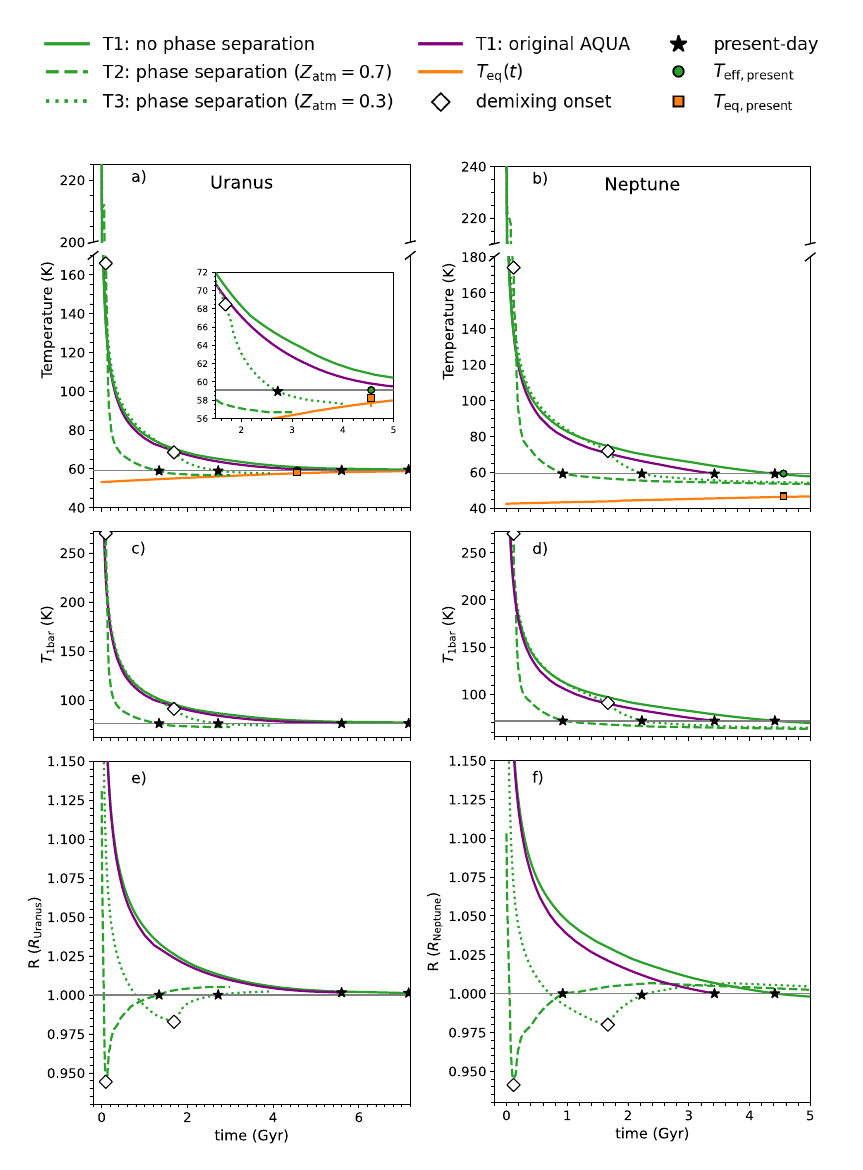}
\caption{Evolution of Uranus (left) and Neptune (right) in terms of effective and equilibrium temperature (a,b), 1-bar temperature (c,d), and  radius (e,f) for the tracks T1 (green solid), T2 (green long-dashed), and T3 (green short-dashed). Tracks T2 and T3 are with phase separation while T1 is without. The time when the present-day structure models occur are marked by a star. White diamonds indicate the onset of demixing. The horizontal grey lines indicate the observed values. The curves stop when $\Teff$ approaches $\Teq$. The purple lines show the T1 models with original AQUA.}
\label{evol-uranus-neptune}
\end{figure*}

\autoref{evol-uranus-neptune} shows the evolution of Uranus in terms of $\Teff$ (panel a), $\Tone$ (c), and radius (e) for the evolution  tracks T1---T3. Numbers given in this Section are summarized in Table \ref{tab:uranus}.

Uranus with a primordial stratification (T1) cools quickly during the first billion year, yet not enough to reach the observed temperature, which is only met after 8.50 Gyr (beyond the figure's time axis).  The radius decreases monotonically. This constant-composition T1 track cannot explain Uranus' observed low luminosity.

Uranus with a high initial $Z_1(0)=0.7$ (T2)  is initially much smaller,  1.16~$\RU$ compared to 1.6~$\RU$ for T1. This comes from a colder initial interior profile due to the high water content in the outer envelope (0.7 vs.~0.14 for T1). Demixing starts early in the evolution, already at 0.11 Gyrs. It starts at +10 K higher $\Tone$ than found in Paper~1 for the same phase diagram and assumptions on composition.  This is because with the revised AQUA EoS, the profile for $\Tone=$271 K is colder than with the original AQUA. At the onset of demixing, the radius has already decreased to 0.936 $\RU$. 
When demixing begins to operate, $\Teff$ decreases rapidly while the radius inflates. This raises the question whether the luminosity  $ L_\mathrm{eff} \sim R_p^2\,\Teff^4$ would increase, because of a larger surface area to radiate from, or whether it would decrease, because of a lower radiation temperature. We find that the luminosity decreases (\autoref{lum-uranus-neptune}a).
With further cooling of the entire planet, demixing progresses and leads to a redistribution of mass in the interior. The decrease in density of the outer envelope with the ongoing loss of water leads to an expansion of the planet. Already at 1.27 Gyr,  $\Teff$ reaches the observed value and the present-day structure model is met. After this point, phase separation proceeds at $\Tone < 76$~K. 

At 2.7 Gyr the Uranus' effective temperature reaches the equilibrium temperature. By then, the radius has grown to 1.005 $\RU$ while $\Tone$ is down to  72~K. 

The Uranus model with demixing and a moderate initial outer envelope abundance of 0.31 (T3) falls in between those of T2 and T1. Its initial radius is 1.445 $\RU$. Demixing sets in at 1.72 Gyr, when the surface temperature has decreased to 91 K. The radius at that point has fallen to 0.981 $\RU$. As for T2, with the onset of demixing the radius begins to increase and $\Teff$ drops. The present-day structure model is reached at 2.68 Gyr. Demixing continues at $\Tone<76\:$K until at 3.32 Gyr  the planet's $\Teff$ reaches $\Teq$.

\subsection{Uranus with equilibrium evolution}

For any considered initial state, $\Teff$ falls for $\Teff>\Teq$ while $\Teq$ increases. In Uranus models with demixing, the initial state has a low entropy, leading to a low energy budget and given Uranus is so close to equilibrium, $\Teff$ reaches $\Teq$ before the present time.  Since the stellar irradiation increases over time and the atmosphere of Uranus is known to be capable of absorbing some sun-light ($A_B<1$), $\Teq$ keeps increasing further. What happens to $\Teff$?

While it is not physically prohibited that $\Teff$ could fall below $\Teq$ as evidenced by Industrial Earth, which now has an energy imbalance  $\sigma_B\, (\Teff^4 -\Teq^4)\simeq-1$~W/m$^2$ \citep{AGU2025}, current Uranus has a small positive net loss ($\Teff\gtrsim \Teq$) and therefore, we assume a state $\Teff \gtrsim \Teq$ applies to Uranus since its first occurrence in time onward.

Since $\Teq$ increases over time, we expect $\Teff$ in equilibrium evolution to increase as well, and possibly $\Tone$. 
It is unclear how equilibrium evolution affects the cooling of the deep interior.  \cite{nettelmann16} assumed that an isothermal region in the deep atmosphere would develop as $\Tone$ begins to increase while the deep interior cools slowly further. It is also possible that demixing ceases because the net heat loss in equilibrium evolution  will be small and thus the change in the deep thermal state be slowed down. 
We simply assume that once the falling $\Teff$ is about to hit the increasing $\Teq$,  Uranus will from then on evolve at  $\Teff\gtrsim \Teq$  until the present day. This means that an evolutionary scenario that reaches equilibrium naturally explain Uranus's observed luminosity. This is the case for scenarios T2 and T3. We caution that it is not obvious that the state of equilibrium evolution can easily be reached. In fact, \citet{scheibe21} found that in the case of early heat retention with later release (in their case due to a decaying TBL), a state with $\Teff$ a few K higher than $\Teq$ would establish and be maintained over billions of years, and thus overestimate the luminosity of present-day Uranus.

\subsection{Uranus with original AQUA}

Uranus  with original AQUA cools faster (6.9 Gyrs) than with Revised AQUA (8.5 Gyrs) when otherwise the same assumptions on composition and bulk structure are made (compare the solid purple and solid green curves in \autoref{evol-uranus-neptune}a). Both curves are for the no-demixing case (T1). The same effect is qualitatively seen for Neptune.

\section{Results for Neptune}
\label{sec:results-neptune}

\begin{table*}
\centering
\caption{Results for evolution tracks for Neptune at different times $t$. }

\label{tab:neptune}
\begin{tabular}{c c c c c c c c c c c c}
\hline
  & $Z_1(0)$ & $Z_2(0)$ & $Z_1(t_{\rm pds})$ & $\Zmeth(t_{\rm pds})$  & $t_{\rm pds}$
  & $t_{\mathrm{sep},i}$ & $\Tone(t_{\mathrm{sep},i})$ & $t_{\mathrm{sep},f}$ & $\Teff(t_{\odot})$ & $R_p(\tau_{\odot})$ & \\
  & & & & & (Gyr) & (Gyr) & (K) & (Gyr) & (K) & ($\RN$)\\
\hline
$T_1$ & 0.10  & 0.905 &  0.10 & 0 &  4.5  & --  & -- & -- & 58.8 & 1.000 \\
$T_2$ & 0.70  & 0.873 &  0.10 & 0 &  0.93 &  0.12 & 271 & 3.24 &  53.6 & 1.003\\
$T_3$ & 0.31  & 0.839 &  0.10 & 0 &  2.23  &  1.67 &  91   & 3.66 &  54.5 & 1.005 \\
$T_4$ & 0.19  & 0.912  &  0.179 & 0.095 & 3.72 &  3.56 &  74  &   6.33 & 56.2 & 1.004  \\
$T_5$ & 0.158  & 0.922 &  0.152 & 0.095 & 3.86 &  3.64 &  74  &  6.81 & 57.3 & 1.006 \\
$T_6$ & 0.179  & 0.922 &  0.179 & 0.095 & 3.97 &  -- &  --  &  -- & 57.9 & 0.996 \\
\hline
\end{tabular}
\end{table*}

The right-hand-side panels of \autoref{evol-uranus-neptune} show the evolution of Neptune 
in terms of $\Teff$, $\Tone$, and radius for tracks T1---T3. Numbers given in this Section are summarized in Table \ref{tab:neptune}.

For Neptune without demixing (T1), the cooling time matches the age of the solar system: at 4.42 Gyr, $\Teff =59.3$, and at 4.56 Gyr, $\Teff=58.8$ K. These values fall within the observational uncertainty  of $\Teff$.
The radius in this scenario decreases monotonically from an initial value of 1.52 $\RN$. 

\begin{figure}
\centering
\includegraphics[width=0.5\textwidth]{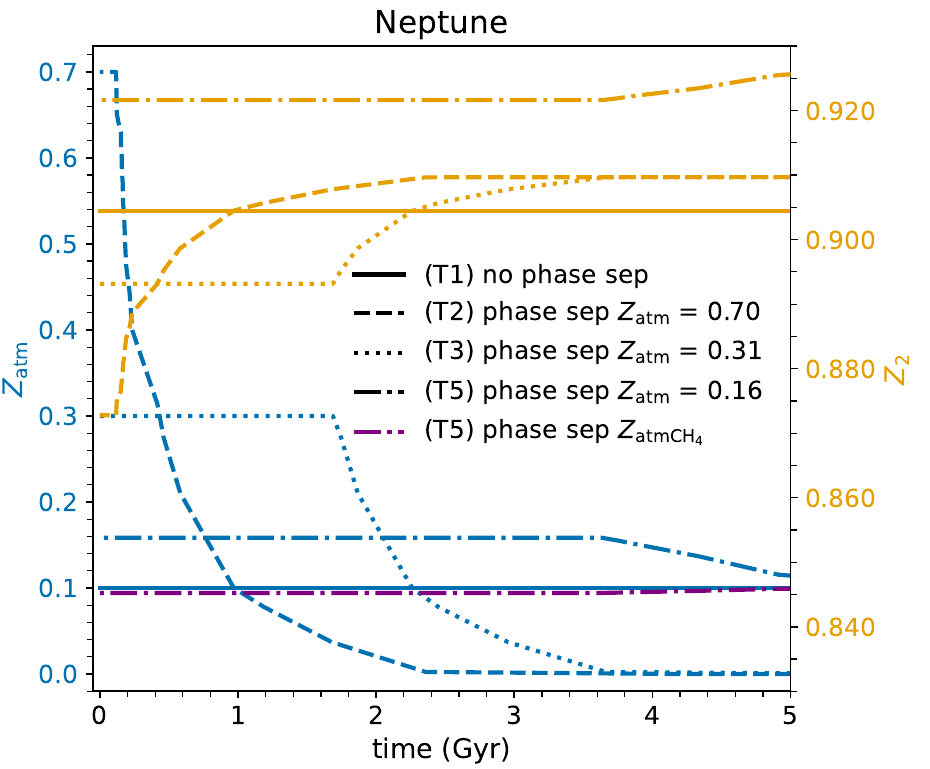}
\caption{Evolution of Neptune's distribution of volatiles in the atmosphere and outer envelope, $Z_{\mathrm{atm}}$ (blue; left axis), and in the inner envelope $Z_{2}$ (orange; right axis). Solid lines are for the T1 model, dashed for T2, dotted  for T3 and dash-dotted for T5. Water abundance changes strongly with the onset of demixing (T2 and T3 curves) and demixing in Neptune proceeds after the present-day structure is reached until complete water depletion of the atmosphere. The purple curve shows the atmospheric methane abundance that increases once demixing starts. The inner envelope methane abundance remains constant at 0.095 and is not shown here. }
\label{neptune-envelopes}
\end{figure}

If the initial water abundance is high (T2), the onset of demixing occurs early, similar to the Uranus-T2 case. The rapid water depletion in the outer envelope (see \autoref{neptune-envelopes}) leads  to rapid increase in radius while $\Teff$ drops. The present-day structure model is reached after only 0.93 Gyrs of cooling. Past this time, demixing continues until no water is left in the outer envelope. This occurs after 3.24 Gyr, when $\Tone$ has fallen to 65 K. At this point the increasing radius has reached 1.005 $\RN$. Once demixing is finished, the planet continues to evolve with a constant composition and the radius decreases again. At 4.56 Gyr, $\Teff=53.6$ K, i.e., 5~K lower than observed, and the radius is 1.003 $\RN$.

For moderate initial water abundance (T3), demixing starts later, at 1.67 Gyr. In the early phase of phase separation, we see again the behavior that the radius increases rapidly while $\Teff$ and $\Tone$ drop. The present-day structure model is reached at 2.23 Gyr, thus again much earlier than in the classical case without compositional evolution (T1). Complete atmospheric water depletion is reached at 3.66 Gyr. At 4.56 Gyr, $\Teff = 54.5$ K and $\Tone=$65 K.

The shorter cooling times of Neptune with phase separation  than without are a result of a cooler initial states that in turn results from a more homogeneous initial planet. Conversely, lower initial $Z_1$-values, which imply  a more  inhomogeneous initial structure, lead to longer cooling times. Since the lowest initial $Z_1$-value occurs for the model without phase separation, the T1-model poses a cap to the maximum possible cooling time under the assumption of an adiabatic, layered structure with water only as the heavy element in the envelopes. 

We find that phase separation provides a potential explanation for Uranus's low present-day luminosity, but it does not provide a consistent explanation for Neptune's brightness. Next we consider possible further contributions to Neptune's thermal evolution.

\subsection{Neptune with methane}

\begin{figure}
\centering
\includegraphics[width=0.49
\textwidth]{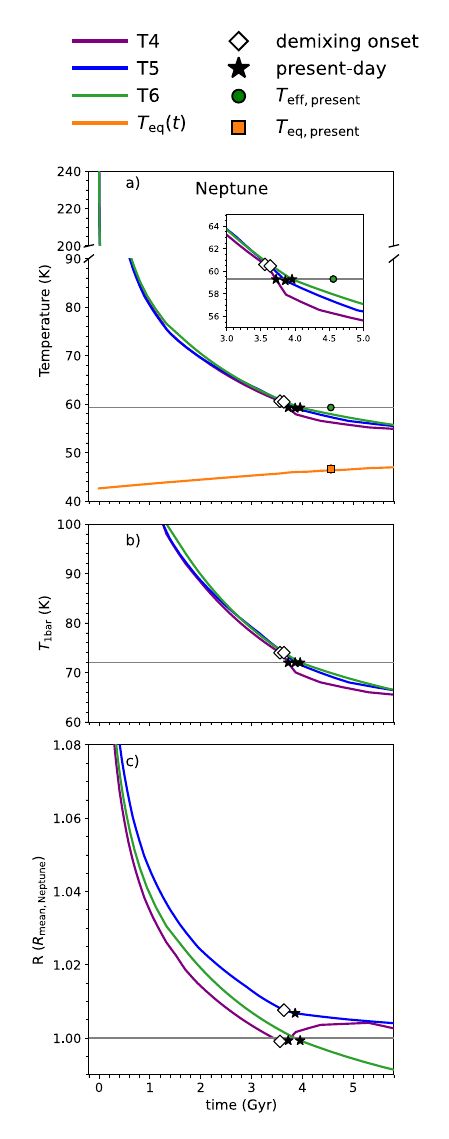}
\caption{Neptune evolution with methane assuming constant methane mass fractions over time (T4, purple) or constant CH$_4$/H$_2$ and He/H particle ratios (T5, blue), or no phase separation (T6, green).
}
\label{fig:neptune-methane}
\end{figure}

\autoref{fig:neptune-methane} shows the thermal evolution of Neptune with methane and hydrogen-water phase separation (T4 and T5), and with  methane but without phase separation (T6).  

In the constant methane mass fraction case T4, the model behaves similar to the T2 and T3 models:  
with the onset of demixing, the radius increases while $\Teff$ drops. 
In the constant particle fraction case T5, the mass fraction of methane in the outer envelope increases over time,  (\autoref{neptune-envelopes}). The associated increase in density balances out to some extent the decrease in density due to the decreasing water content. As a result, radius inflation does not occur and the planet contracts continuously. 

Note that the radius of the present-day T5 model is slightly too large (by 0.7\%) .
This is because $\Zwater$ was kept constant at 0.10 when $\Zmeth$ was fitted to the gravitational harmonics, and only afterwards -- instead of simultaneously -- adjusted to the resulting value of $\Zmeth$ under conservation of $x_{\rm H2O}$ (see Appendix \ref{apx:meth_eqs}). The lower $\Zwater=0.057$ in the full mixture causes the slightly larger radius (see Table \ref{tab:neptune}).  

Including methane in the interior models seems to delay the onset of demixing and the cooling of Neptune; phase separation begins only when the planet has already cooled down to 74 K and the cooling time to reach the present-day structure extends to nearly 4 Gyrs.

The slower cooling of models with methane results from three overlapping properties. First, we find that adiabats with a given water mass fraction are generally cooler with methane than without, that is, replacing H/He with methane makes the adiabat colder. This behavior applies to pressures $\gtrsim 0.3$ GPa and thus to pressures around the demixing region. At pressures lower than 0.3 GPa and thus lower temperatures $T\lesssim$1000~K,  the dense, molecular  hydrogen fluid, which does have internal degrees of freedom like rotation and vibration, leads to lower temperatures than in the ideal gas approximation for methane. If the adiabat is colder in the region around 11 GPa, a stronger reduction in the water abundance is needed for consistency with the phase diagram. Therefore, $\Zwater$ of the present-day model is lower in the presence of methane, see Table \ref{tab:neptune}.
From this property of a colder adiabat with methane, one would expect an earlier onset of demixing and a faster cooling, contrary to what we observe.

Second, we find that adiabats with a given total metallicity are generally warmer with methane, that is, replacing water with methane makes the adiabat warmer. For instance, the T4 model has an initial structure with  $\Zmeth=\Zwater=0.095$, and demixing begins at $\Tone=74$ K. We checked that for the same total mass fraction of 0.19 but considering only water, the adiabats become colder and demixing starts earlier, at 80 K.  From this property alone one would expect a delayed onset of demixing with methane and a slower cooling, which is what we observe.
However, we face the case that the amount of water that is replaced by methane is quite small, 0.057 vs.~0.100. This is smaller than the amount of H/He that is replaced by methane (total $Z_1$ of 0.179 vs.~0.100). Thus one would  expect the first property to dominate, which is not what we observe. 

Yet third, lo and behold, we put a cap to the initial water abundance in the outer envelope in the models with methane, which is $\Zwater \leq \Zmeth$, as we assumed C and O were accreted in form of CO ice on top of a heavy-element-rich core, whatever it is exactly made of. As it turns out now, this is a strong constraints.  Since we require a low present methane mass fraction to fit the gravitational moments, the initial water mass fraction must also have been low (we find 0.064), which leads to solutions closer to the T1 track ($\Zwater=0.1$) than the T2, T3 tracks (initial $\Zwater > 0.3$).

Track T5 behaves similar to T4 except for the monotonic shrinking. This is our favorite scenario as constant particle ratio with respect to hydrogen when only water rains out is physically motivated, while the T4 scenario is merely illustrative. The T5 track has a cooling time of 3.86~Gyrs.

\subsection{Neptune with latent heat}
\label{sec:nep_latent}

So far we neglected the effect of latent heat occurring upon water phase transitions. Here, we investigate this effect on the evolution of the T1-model and compare the $P-T$ profiles of that model with and without an entropy jump that leads to the latent heat release. 

\paragraph{Method}
As the planet cools, a growing region of the interior crosses the superionic (SI) phase boundary, and latent heat is released.
At each time step $t_i$, we locate the mass coordinate $m_{\rm SI}$ at which the adiabat crosses the fluid-superionic phase boundary $T_\mathrm{SI}(P)$ and determine the entropy jump $\Delta s_\mathrm{SI}(P,T)$ across. Underneath, the temperature profile follows the lower-entropy isentrope. 

For each pair of successive time steps $t_i$, $t_{i+1}$ we evaluate the mass in the planet
$ dm_{\mathrm{SI}} = m_{\mathrm{SI}}(t_{i+1}) - m_{\mathrm{SI}}(t_i) $
within which the new conversion to the superionic phase occurs. The additional luminosity due to the latent heat is approximated as 
\begin{equation}\label{eq:Llatent}
    L_{\mathrm{lat}} = \delta\dot{Q}_\mathrm{latent} \approx \displaystyle\frac{1}{2\:dt}\left[T(t_i)\, \Delta s_\mathrm{SI}(t_i) + T_{i+1}\, \Delta s_\mathrm{SI}(t_{i+1})\right]\, Z_2\, dm_\mathrm{SI}.
\end{equation}
The additional source term $L_{\mathrm{lat}}$ must not be included in the energy balance equation (\ref{eq:Leff}) if the latent heat is already accounted for by an entropy jump in the $P$--$T$ profile as we do here.

\paragraph{Result}

\begin{figure*}
\centering
\includegraphics[width=1.0
\textwidth]{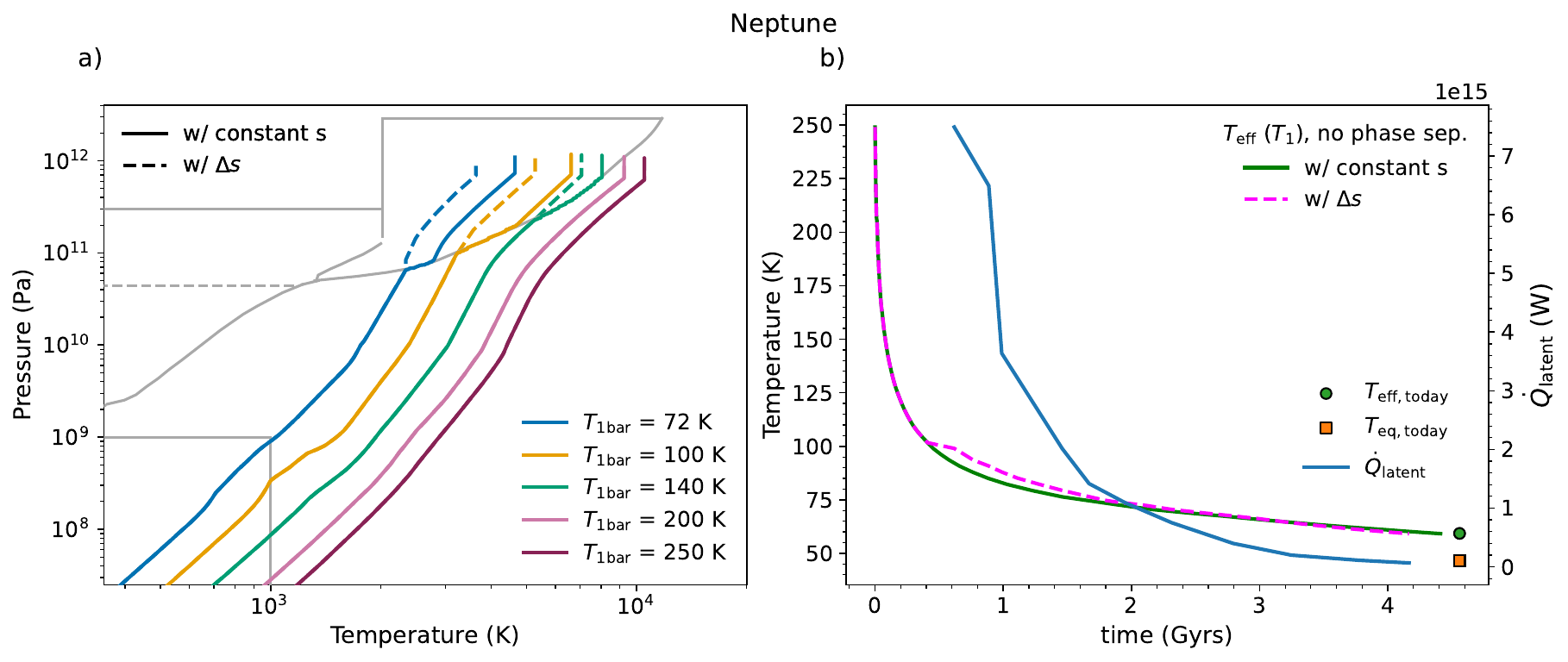}
\caption{Effect of latent heat release from the liquid-to-superionic water phase transition on Neptune. a) adiabats of the track T1- model at selected $\Tone$ values of 72, 100, 140, 200, and 250 K. The boundaries of the water phase diagram (grey) are adapted from \citet{canoamoros26}. The solid (constant entropy) and dashed (with entropy jump) curves differ when the latter cross the liquid-superionic boundary, after which the latter follow a lower entropy adiabat. b) Resulting thermal evolution for track T1 without (green, solid) and with latent heat (magenta, dashed). The blue curve shows the rate of latent heat release $\dot{Q}$.}
\label{LH panel}
\end{figure*}

Above $\Tone =200$ K, the adiabats stay entirely within the fluid phase (\autoref{LH panel}a). At $\Tone = $ 140 K, a small superionic layer has begun to form above the core. It grows outward. At $\Tone = $ 100 K, a substantial fraction of the planet has crossed into the superionic region, and the $P$--$T$ profiles in \autoref{LH panel}a with entropy jump (dashed lines) differ from those obtained using constant entropy (solid lines) by up to almost 1000 K. The latter continue along the phase boundary, while the $P-T$ profiles with entropy jump follow the lower-entropy superionic adiabat into the deeper interior and become colder.

\autoref{LH panel}b shows the influence of including the latent effect $T\, \Delta s_{\rm SI}\,dm_{\rm SI}$ vs.~neglecting it (the T1-track) on the thermal evolution. As Neptune cools past $\Tone = $ 140 K, the evolution curves begin to separate. Neptune with latent heat release cools at the beginning more slowly. The blue curve shows the rate of latent heat release, $\dot{Q}$. It peaks early and decays as the amount of newly "freezing" mass decreases. After a billion-year-long brighter epoch, the planetary interior has lost more of its initial energy budget so that at present time it appears fainter. Neptune's cooling time with the latent heat effect shortens by $-0.3$~Gyr to 4.2 Gyrs.

\section{Discussion}
\label{sec_disc}

\subsection{A primordial Neptune?}
\label{sec:disc_adN}

For the Neptune models with a primordial composition and no methane, the use of Revised AQUA compared to original AQUA prolongs the cooling time by $\sim 1$ Gyr, which is just about enough to meet Neptune's observed luminosity at the present time. 

While we cannot trace this strong finding uniquely back to the properties of revised compared to original AQUA, there are some clear differences that likely play a role.
The revision led to a $P-T$ dependent shift in the entropy, as can be seen in Figure 3 of \citet{canoamoros26}. As the planet cools, the adiabats move through a non-uniform entropy difference between the two EoSs. 

The revised adiabats profiles have a higher specific entropy in the inner envelope than the original profiles at the same $\Tone$. The difference in entropy also varies with $\Tone$. The curves start diverging at around $\Tone = 260$ K. Below that, the revised EoS yields a larger energy change per step due both to a larger entropy change and a larger mean temperature. Radiating away a larger amount of energy at the same atmospheric conditions takes longer, leading to the slower cooling.

Our result a of cooling time of Neptune of $\sim$4.5 Gyrs when a water EoS is used as a proxy for heavy elements adds to a line of successive improvements in the EoSs. \citet{HMacF80} used a fit formula to theoretical prediction constrained by experimental data up to $\sim$2 Mbar for ices. They obtained too long cooling times for Neptune. \citet{FN09} applied improved H/He- and water EoSs. Both with the EoSs from the SESAME database \citep{SESAME} and with H/He-REOS.1 and H2O-REOS, the luminosity of Neptune was explained by a primordial three-layer structure that would adiabatically cool, like we find here.  Further improvements in the H/He EoS, be it H/He-REOS.3 \citep{Becker14} or CD21 led to a faster cooling \citep{scheibe19, HowardHelledAA24}. While the accelerated cooling was primarily not due to the water EoS used, the water EoS used by \citet{scheibe19} was that by \citet{Mazevet2019}, and that EoS is the one that is incorporated into original AQUA for the dense fluid and superionic region. This region is most relevant for the ice giants, and is found in this work to lead to the $-1$~Gyr faster cooling with original AQUA.

A substantial improvement in that area of phase space was made by \citet{French2016} who provided the entropy of superionic water. These data are included in Revised  AQUA. This study resolves a first-order fluid-to-superionic phase transition in water, associated with a latent heat of $\sim$1\ $\mathrm{J\,g^{-1}\,K^{-1}}$. However, for our evolution runs we ignored this phase transition, and assumed constant entropy across instead. This assumption is made because the superionic phase may not form under the conditions in the real planet as compared to the idealized conditions of a H/He/water mixture. Instead, it is possible that the presence of impurities such as silicates delay or suppress this phase transition. For instance, \citet{Darafeyeu24} find a strong shift of the melting line of the superionic phase in their DFT-MD simulations of 1:4 SiO$_2$--H$_2$O mixtures compared to that of pure water so that even present-day Neptune would be too warm to freeze. The T1 model for Neptune with revised AQUA has a core temperature of 4340~K in excellent agreement with the A-type Neptune model of \citet{Militzer_2025} who obtains 4300~K.

Our finding that with state-of-the-art EoSs for H/He and water the luminosity of Neptune is recovered can still be a coincidence --as it was before. Taken at face value, the agreement would suggest a layered structure with compositional gradients that are too steep to be eroded over time, don't inhibit heat transport, and arise from factors other than phase separation, the delayed or absent freezing of superionic water, and a negligible influence of methane.   

Adding methane destroys the picture of a primordially-structured Neptune. If methane were just to replace water, the adiabat would become warmer and the cooling time longer, leaving more flexibility on other assumptions. However, the total abundance of the ices that is necessary to match the present-day observational constraints rises when methane is added, and overall the adiabats become cooler and thus the cooling times shortens to below the age of the Solar System. The influence of the assumed C:O ratio on the evolution of icy adiabatic Uranus and Neptune will be studied in future work.

\subsection{A reflective Uranus?}
\label{sec:disc_albU}

The Bond albedo determines the fraction of incident stellar radiation reflected by the planet and therefore affects the absorbed stellar flux entering the planet's energy budget. Bond albedo estimates are subject to observational uncertainties and may change as improved measurements become available. For example, Jupiter's estimated Bond albedo was revised from 0.343, based primarily on Pioneer and Voyager 1 observations \citep{TOMASKO1978,Hanel1981}, to $0.503 \pm 0.012$ using Cassini observations \citep{Li2018}. Similarly, a recent reassessment of Uranus' energy balance by \citet{irwin2025} derived an orbital-mean Bond albedo of $0.349 \pm 0.016$ with seasonal variations between 0.335 and 0.363, higher than the 0.3 measured by the Voyager 2 spacecraft \citep{pearlconrath1991}.
\citet{scheibe19} showed that the response of Uranus' cooling time is sensitive to the assumed albedo value. They found an increase to 0.4, sufficient to explain the luminosity for their T1-like adiabatic evolution models and adopted EoS.   

\begin{figure}
\centering
\includegraphics[width=0.5\textwidth]{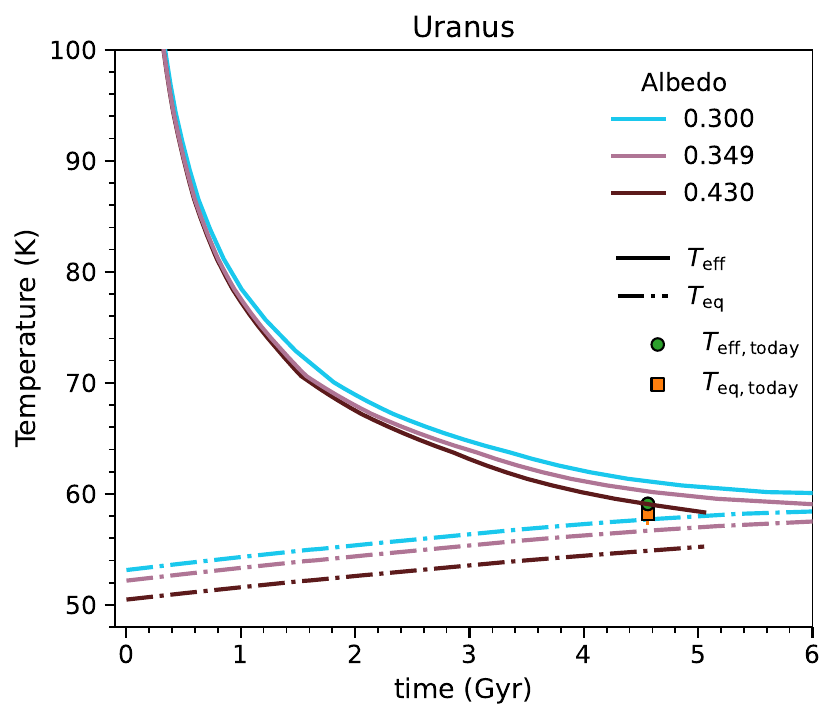}
\caption{Thermal evolution of Uranus Track 1 (constant composition) for three assumed values of the albedo. Solid and dash-dotted curves show the effective temperature, $T_{\mathrm{eff}}$, and equilibrium temperature, $T_{\mathrm{eq}}$, respectively. Colours denote albedoes of 0.300 (blue), 0.349 (pink), and 0.430 (maroon). The value 0.300 is the measured albedo reported by \citet{PEARL1990}, while 0.349 is the recent estimate of \citet{irwin2025}. The value 0.430 was obtained iteratively to reproduce the observed luminosity. The green circle and orange square indicate the observed $T_{\mathrm{eff}} = 59.1 \pm 0.3$~K, and $T_{\mathrm{eq}} = 58.2 \pm 1.0$~K, respectively.}
\label{albedo-uranus}
\end{figure}

Here, using different EoSs, we ask again how much higher than the nominal Voyager value the albedo would have to be for adiabatic T1-like evolution models to explain the luminosity of Uranus. The answer is $A = 0.43$ (Figure \ref{albedo-uranus}). While this is higher than the observed by 0.13 and by 0.08 with respect to the reassessed Voyager value, the enhancement is smaller than the  enhancement by $+$0.16 measured by Cassini for Jupiter. We advocate to aim for an accurate Bond Albedo measurement in a future mission to the Uranian system.   

\subsection{Uranus and Neptune with demixing in comparison}

Can the different present-day luminosities of Uranus and Neptune be explained simply by assuming that they underwent different amounts of H$_2$--H$_2$O phase separation just to end up with similar interior structures and compositions today? With moderate to high ($\sim$0.3---0.7) initial outer envelope water abundance, both planets have initially a low energy budget and become faint early. Uranus undergoes equilibrium evolution, Neptune does not. Because of that, Uranus cools very slowly today and therefore, demixing proceeds very slowly and would release only little --if at all-- gravitational energy from this process -- as envisioned by \citet{Bailey_2021}. However, they suggested complete demixing has occurred for Uranus while we find equilibrium evolution, which actually prevents Uranus from complete demixing. In contrast, we find that Neptune can undergo complete demixing. Neptune with demixing becomes too faint today though and requires an additional heat source to explain its brightness. We conclude that Uranus and Neptune cannot simply be explained by different levels of demixing.

\subsection{An additional heat source in Neptune?}

According to our results, Neptune with phase separation requires an additional energy source to explain its observed brightness. A too faint Neptune-sized planet at the age of the solar system was also obtained by \citet{HowardHelledAA24} unless they enhanced the demixing temperatures with respect to their version of the phase diagram of \citet{berg24}, in which case Neptune would not become too faint but too big.

Any of the processes studied previously for Uranus and Neptune without phase separation might play a role
in powering Neptune's luminosity. Previous suggestions range from a deep TBL at a sharp ice-poor/heavy-element-rich transition that builds up (Uranus case) or decays (Neptune case), a deep TBL alongside near-equilibrium evolution where $\Teff$ stays over billions of years at about the observed value, which is obtained as a natural outcome of a consistent treatment of heat flux and temperature gradient \citep{scheibe21}, an extended deep $Z$-gradient \citep{hubbard1995, vazan20, tejada-arevalo2025}, a superionic freezing core at the top of which a TBL develops as a result of sluggish convection in the viscous core \citep{stix21}. To all these processes, cloud condensation can also contribute \citep{MarkhamStevenson21}.

Gravitational energy release from phase separation between other constituents, such as within the system of planetary ices \citep{Militzer24} or the C-H system \citep{Kraus17}, could also be the missing energy source.  Given our finding that demixing in the O-H system lets Neptune appear even fainter, it is open in which direction demixing in other systems would drive the luminosity of Neptune.

\subsection{Future direction}

There are many directions for improving our understanding of Uranus and Neptune as the ice giant prototypes. We address a few of them.

Our models rely on the Revised AQUA EoS for water, a compiled methane EoS, and our version of the H$_2$O-H$_2$ phase diagram. Those are certainly not the last answers as ongoing efforts in improving the water Eos \citep{Militzer25} and the understanding of phase separation among mixtures of volatiles  \citep{Militzer24} suggest. As a first step it would be important to obtain experimental constraints on the phase diagram at pressures of 5 to a few tens of GPa  since this is the region where interior models predict a compositional change.

Rocks may be distributed within the inner envelope rather than confined into a central core. While a homogeneous distribution within a volatile-rich inner envelope would not affect the evolution much, a compositional gradient retarding heat loss would. Experiments find miscibility of SiO$_2$ in water at $\sim$30--50 GPa \citep{Nisr20} and of MgO in water at 10--30 GPa \citep{Stalder01,Kim21} at temperatures below the Neptune adiabat, but  numerical simulations suggest miscibility of MgO and water at higher pressures to occur at higher temperatures than along the Neptune adiabat \citep{Kovacevic23}, implying ice-rock immiscibility in the deep interior.  How that influences the coupled structure and evolution remains to be investigated.   

The temperature where the adiabat begins is very important for processes deeper inside. While we have assumed the adiabat to begin at 1 bar, water clouds at deeper levels may significantly change the deeper temperature profile due to latent heat release \citep{Kurosaki17}, leading to a colder interior, or due to the suppression of convection \citep{Leconte17}, leading to a warmer interior. Future work should include the influence of water clouds in dependence of the water abundance.

The C-H system may be of no lesser importance. Indeed, diamond formation due to the phase separation in carbohydrates like methane, cane sugar, or plastics is now  routinely seen in lab experiments \citep{he22}. Diamond rain in the ice giants is a candidate energy provider and should be included in future evolution models of the ice giants. 

This work lines-up with numerous previous recommendations that a future space mission to an ice giant should carry an entry probe for measuring the atmospheric abundances of oxygen, carbon, and the noble gases. If C and O were accreted as CO ice as assumed in our Neptune models with methane, the He/H and Ne/H ratios could be higher than protosolar. Or they could be lower if hydrogen was released from phase separation in the C-H system at greater depths.

\section{Conclusion}
\label{sec_conc}
\begin{enumerate}
\item 
The application of the revised AQUA EoS \citep{canoamoros26} prolongs the cooling time of adiabatic, layered structure models (T1) compared to original AQUA \citep{aqua2020} and earlier water EoS. 
\item
For Uranus, the application of revised AQUA on T1-type models leads to a too slow cooling unless the Bond Albedo is $\sim$0.43 instead of the observed $\sim$ 0.3: we recover the faintness-problem of Uranus.
\item 
For Neptune, the application of revised AQUA together with the modern CD21 EoS for H and He in T1-type models explains its luminosity, provided that the methane abundance in the outer envelope is much lower than observed in the atmosphere, and that water in the deep interior began freezing to the superionic phase only recently, if at all: the recently emerged brightness-problem of Neptune \citep{scheibe19} seems to disappear.  
\item 
Phase separation of hydrogen and water would influence the cooling behavior of Uranus and Neptune strongly even if they contained only 30\% water by mass. 
Assuming that water is the only heavy element in the outer envelope (tracks T2 and T3), the onset of phase separation leads to a pronounced increase in planet radius and a decrease in luminosity. This is because the energy gain from cooling goes into the gravitational energy for expanding the planet rather than being radiated away. We do not observe a noticeable effect of the interior gravitational energy gain from sinking water droplets.  
\item 
For Uranus with phase separation, $\Teff$ decreases until $\Teq$ is met, after which the planet could evolve in equilibrium with the incident flux until the present-day. There is a wide range of initial conditions and envelope water contents for which the current luminosity could be obtained, provided that equilibrium evolution, which we not model explicitly, is a viable path. 
\item 
For Neptune with phase separation, an additional energy source is required. An obvious follow-up project is to study demixing among other abundant constituents such as in the C-H system to determine whether they would provide that energy or else like we find for the H-O system, act in the opposite direction, and how Uranus would respond. 
\end{enumerate}

\section*{Data availability}
The methane EoS table can be downloaded from the Figshare repository at 
\url{https://doi.org/10.6084/m9.figshare.34845720} .

\begin{acknowledgements}
      We thank M.~Bethkenhagen for discussions on the phases of methane. MCA and NT acknowledge the support of the German Science Foundation (DFG) through the research unit FOR 2440 “Matter under planetary interior conditions” (projects PA 3689/1-1 and NE~1734/2-2). NN and MCA acknowledge support through DFG-grant NE~1734/3-1.
      
\end{acknowledgements}

\bibliographystyle{aa} 
\bibliography{bibliography}

@article{MacKenzie_2023,
	author = {MacKenzie, Jasmine and Grenfell, {John Lee} and Baumeister, Philipp and Tosi, Nicola and Cabrera, Juan and Rauer, Heike},
	title = {Effect of improved atmospheric opacities in modelling sub-Neptunes - A case study for GJ 1214 b},
	DOI= "10.1051/0004-6361/202141784",
	url= "https://doi.org/10.1051/0004-6361/202141784",
	journal = {A\&A},
	year = 2023,
	volume = 671,
	pages = "A65",
}

@article{Baumeister_2023,
	author = {Baumeister, Philipp and Tosi, Nicola},
	title = {ExoMDN: Rapid characterization of exoplanet interior structures with mixture density networks★★★},
	DOI= "10.1051/0004-6361/202346216",
	url= "https://doi.org/10.1051/0004-6361/202346216",
	journal = {A\&A},
	year = 2023,
	volume = 676,
	pages = "A106",
}

@article{lindal1987,
  title={The atmosphere of Uranus: Results of radio occultation measurements with Voyager 2},
  author={Gunnar F. Lindal and James R. Lyons and Donald Sweetnam and Von R. Eshleman and David Hinson and G. Leonard Tyler},
  journal={Journal of Geophysical Research},
  year={1987},
  volume={92},
  pages={14987-15001},
  url={https://api.semanticscholar.org/CorpusID:120057515}
}

@article{Becker14,
   title={AB INITIO EQUATIONS OF STATE FOR HYDROGEN (H-REOS.3) AND HELIUM (He-REOS.3) AND THEIR IMPLICATIONS FOR THE INTERIOR OF BROWN DWARFS},
   volume={215},
   ISSN={1538-4365},
   url={http://dx.doi.org/10.1088/0067-0049/215/2/21},
   DOI={10.1088/0067-0049/215/2/21},
   number={2},
   journal={The Astrophysical Journal Supplement Series},
   publisher={American Astronomical Society},
   author={Becker, Andreas and Lorenzen, Winfried and Fortney, Jonathan J. and Nettelmann, Nadine and Schöttler, Manuel and Redmer, Ronald},
   year={2014},
   month=dec, pages={21} }

@article{bali13,
  author =       "{Bali}, E and {Audétat}, A and {Keppler}, H.",
  title =        "Water and hydrogen are immiscible in Earth’s mantle",
  journal =      "Nature",
  volume =       "495",
  year =         "2013",
  DOI =          "https://doi.org/10.1038/nature11908"
}

@article{Bailey_2021,
doi = {10.3847/PSJ/abd1e0},
url = {https://dx.doi.org/10.3847/PSJ/abd1e0},
year = {2021},
month = {3},
publisher = {The American Astronomical Society},
volume = {2},
number = {2},
pages = {64},
author = {Elizabeth Bailey and David J. Stevenson},
title = {Thermodynamically Governed Interior Models of Uranus and Neptune},
journal = {The Planetary Science Journal}
}

@article{Baumeister_2020,
doi = {10.3847/1538-4357/ab5d32},
url = {https://dx.doi.org/10.3847/1538-4357/ab5d32},
year = {2020},
month = {1},
publisher = {The American Astronomical Society},
volume = {889},
number = {1},
pages = {42},
author = {Philipp Baumeister and Sebastiano Padovan and Nicola Tosi and Grégoire Montavon and Nadine Nettelmann and Jasmine MacKenzie and Mareike Godolt},
title = {Machine-learning Inference of the Interior Structure of Low-mass Exoplanets},
journal = {The Astrophysical Journal}
}

@ARTICLE{jacobson2009,
       author = {{Jacobson}, R.~A.},
        title = "{The Orbits of the Neptunian Satellites and the Orientation of the Pole of Neptune}",
      journal = {AJ},
         year = 2009,
        month = may,
       volume = {137},
       number = {5},
        pages = {4322-4329},
          doi = {10.1088/0004-6256/137/5/4322},
       adsurl = {https://ui.adsabs.harvard.edu/abs/2009AJ....137.4322J}
}

@article{HowardHelledAA24,
	author = {Howard, S. and Müller, S. and Helled, R.},
	title = {Evolution of Jupiter and Saturn with helium rain},
	DOI= "10.1051/0004-6361/202450629",
	url= "https://doi.org/10.1051/0004-6361/202450629",
	journal = {A\&A},
	year = 2024,
	volume = 689,
	pages = "A15",
}

@article{MarkhamStevenson21,
doi = {10.3847/PSJ/ac091d},
url = {https://dx.doi.org/10.3847/PSJ/ac091d},
year = {2021},
month = {aug},
publisher = {The American Astronomical Society},
volume = {2},
number = {4},
pages = {146},
author = {Steve Markham and Dave Stevenson},
title = {Constraining the Effect of Convective Inhibition on the Thermal Evolution of Uranus and Neptune},
journal = {The Planetary Science Journal},
}

@article{Vlasov23,
       author = {{Vlasov}, Kirill and {Aud{\'e}tat}, Andreas and {Keppler}, Hans},
        title = "{H$_{2}$-H$_{2}$O immiscibility in Earth's upper mantle}",
      journal = {Contributions to Mineralogy and Petrology},
         year = {2023},
        month = {7},
       volume = {178},
       number = {7},
          eid = {36},
        pages = {36},
          doi = {10.1007/s00410-023-02019-7},
       adsurl = {https://ui.adsabs.harvard.edu/abs/2023CoMP..178...36V}
}

@article{Nettelmann11,
         author = {Nettelmann, N. and Fortney, J.J. and Kramm, U. and Redmer, R.},
         year = {2011},
         Journal={ApJ},
         Volume={733}, Pages={2},
         Title = {Thermal evolution and structure models of the transiting super-Earth GJ1214b}
         }

@article{berg24,
  title = {Ab initio calculation of the miscibility diagram for mixtures of hydrogen and water},
  author = {Bergermann, Armin and French, Martin and Redmer, Ronald},
  journal = {Phys. Rev. B},
  volume = {109},
  pages = {174107},
  numpages = {6},
  year = {2024},
  doi = {10.1103/PhysRevB.109.174107},
}

@article{nett2013,
title = {New indication for a dichotomy in the interior structure of Uranus and Neptune from the application of modified shape and rotation data},
journal = {Planetary and Space Science},
volume = {77},
pages = {143-151},
year = {2013},
issn = {0032-0633},
doi = {https://doi.org/10.1016/j.pss.2012.06.019},
url = {https://www.sciencedirect.com/science/article/pii/S0032063312001900},
author = {N. Nettelmann and R. Helled and J.J. Fortney and R. Redmer},
}

@article{scheibe19,
	author = {Scheibe, Ludwig and Nettelmann, Nadine and Redmer, Ronald},
	title = {Thermal evolution of Uranus and Neptune - I. Adiabatic models},
	DOI= "10.1051/0004-6361/201936378",
	url= "https://doi.org/10.1051/0004-6361/201936378",
	journal = {A\&A},
	year = 2019,
	volume = 632,
	pages = "A70",
}

@article{scheibe21,
   title={Thermal evolution of Uranus and Neptune: II. Deep thermal boundary layer},
   volume={650},
   ISSN={1432-0746},
   url={http://dx.doi.org/10.1051/0004-6361/202140663},
   DOI={10.1051/0004-6361/202140663},
   journal={A\&A},
   publisher={EDP Sciences},
   author={Scheibe, Ludwig and Nettelmann, Nadine and Redmer, Ronald},
   year={2021},
   month=jun, pages={A200} }

@article{nettelmann16,
   title={Uranus evolution models with simple thermal boundary layers},
   volume={275},
   ISSN={0019-1035},
   url={http://dx.doi.org/10.1016/j.icarus.2016.04.008},
   DOI={10.1016/j.icarus.2016.04.008},
   journal={Icarus},
   publisher={Elsevier BV},
   author={Nettelmann, N. and Wang, K. and Fortney, J.J. and Hamel, S. and Yellamilli, S. and Bethkenhagen, M. and Redmer, R.},
   year={2016},
   month=sep, pages={107–116} }

@article{stix21,
doi = {10.3847/PSJ/ac2a47},
url = {https://dx.doi.org/10.3847/PSJ/ac2a47},
year = {2021},
month = {nov},
publisher = {The American Astronomical Society},
volume = {2},
number = {6},
pages = {222},
author = {Lars Stixrude and Stefano Baroni and Federico Grasselli},
title = {Thermal and Tidal Evolution of Uranus with a Growing Frozen Core},
journal = {The Planetary Science Journal},
}

@ARTICLE{vazan20,
       author = {{Vazan}, Allona and {Helled}, Ravit},
        title = "{Explaining the low luminosity of Uranus: a self-consistent thermal and structural evolution}",
      journal = {aap},
         year = 2020,
        month = jan,
       volume = {633},
          eid = {A50},
        pages = {A50},
          doi = {10.1051/0004-6361/201936588},
archivePrefix = {arXiv},
       eprint = {1908.10682},
 primaryClass = {astro-ph.EP},
       adsurl = {https://ui.adsabs.harvard.edu/abs/2020A&A...633A..50V}
}

@article{cd21,
   title={A New Equation of State for Dense Hydrogen–Helium Mixtures. II. Taking into Account Hydrogen–Helium Interactions},
   volume={917},
   ISSN={1538-4357},
   url={http://dx.doi.org/10.3847/1538-4357/abfc48},
   DOI={10.3847/1538-4357/abfc48},
   number={1},
   journal={The Astrophysical Journal},
   publisher={American Astronomical Society},
   author={Chabrier, Gilles and Debras, Florian},
   year={2021},
   month=aug, pages={4} }

@article{aqua2020,
	author = {Haldemann, Jonas and Alibert, Yann and Mordasini, Christoph and Benz, Willy},
	title = {AQUA: a collection of H2O equations of state for planetary models★},
	DOI= "10.1051/0004-6361/202038367",
	url= "https://doi.org/10.1051/0004-6361/202038367",
	journal = {A\&A},
	year = 2020,
	volume = 643,
	pages = "A105",
}

@article{hubbard1989,
title = {Optimized Jupiter, Saturn, and Uranus interior models},
journal = {Icarus},
volume = {78},
number = {1},
pages = {102-118},
year = {1989},
issn = {0019-1035},
doi = {https://doi.org/10.1016/0019-1035(89)90072-9},
url = {https://www.sciencedirect.com/science/article/pii/0019103589900729},
author = {W.B. Hubbard and Mark S. Marley},
}

@article{chabrier19,
   title={A New Equation of State for Dense Hydrogen–Helium Mixtures},
   volume={872},
   ISSN={1538-4357},
   url={http://dx.doi.org/10.3847/1538-4357/aaf99f},
   DOI={10.3847/1538-4357/aaf99f},
   number={1},
   journal={The Astrophysical Journal},
   publisher={American Astronomical Society},
   author={Chabrier, G. and Mazevet, S. and Soubiran, F.},
   year={2019},
   month=feb, pages={51} }

@article{malamud24,
title = {Uranus and Neptune as methane planets: Producing icy giants from refractory planetesimals},
journal = {Icarus},
volume = {421},
pages = {116217},
year = {2024},
issn = {0019-1035},
doi = {https://doi.org/10.1016/j.icarus.2024.116217},
url = {https://www.sciencedirect.com/science/article/pii/S001910352400277X},
author = {Uri Malamud and Morris Podolak and Joshua I. Podolak and Peter H. Bodenheimer},
}

@BOOK{zharkov78,
       author = {{Zharkov}, V.~N. and {Trubitsyn}, V.~P.},
        title = "{Physics of planetary interiors}",
        year = 1978,
        publisher = "Pachart Publishing House",
       adsurl = {https://ui.adsabs.harvard.edu/abs/1978ppi..book.....Z}
}

@ARTICLE{french2024,
       author = {{French}, Richard G. and {Hedman}, Matthew M. and {Nicholson}, Philip D. and {Longaretti}, Pierre-Yves and {McGhee-French}, Colleen A.},
        title = "{The Uranus system from occultation observations (1977-2006): Rings, pole direction, gravity field, and masses of Cressida, Cordelia, and Ophelia}",
      journal = {\icarus},
         year = 2024,
        month = mar,
       volume = {411},
          eid = {115957},
        pages = {115957},
          doi = {10.1016/j.icarus.2024.115957},
archivePrefix = {arXiv},
       eprint = {2401.04634},
 primaryClass = {astro-ph.EP},
       adsurl = {https://ui.adsabs.harvard.edu/abs/2024Icar..41115957F}
}

@article{pearlconrath1991,
author = {Pearl, J. C. and Conrath, B. J.},
title = {The albedo, effective temperature, and energy balance of Neptune, as determined from Voyager data},
journal = {Journal of Geophysical Research: Space Physics},
volume = {96},
number = {S01},
pages = {18921-18930},
doi = {https://doi.org/10.1029/91JA01087},
url = {https://agupubs.onlinelibrary.wiley.com/doi/abs/10.1029/91JA01087},
eprint = {https://agupubs.onlinelibrary.wiley.com/doi/pdf/10.1029/91JA01087},
year = {1991}
}

@article{Leconte17,
	author = {Leconte, Jérémy and Selsis, Franck and Hersant, Franck and Guillot, Tristan},
	title = {Condensation-inhibited convection in hydrogen-rich atmospheres  - Stability against double-diffusive processes   and thermal profiles for Jupiter, Saturn, Uranus, and Neptune },
	DOI= "10.1051/0004-6361/201629140",
	url= "https://doi.org/10.1051/0004-6361/201629140",
	journal = {A\&A},
	year = 2017,
	volume = 598,
	pages = "A98",
}

@ARTICLE{Guillo95,
       author = {{Guillot}, Tristan},
        title = "{Condensation of Methane, Ammonia, and Water and the Inhibition of Convection in Giant Planets}",
      journal = {Science},
         year = 1995,
        month = sep,
       volume = {269},
       number = {5231},
        pages = {1697-1699},
          doi = {10.1126/science.7569896},
       adsurl = {https://ui.adsabs.harvard.edu/abs/1995Sci...269.1697G}
}

@ARTICLE{lindal1992,
       author = {{Lindal}, Gunnar F.},
        title = "{The Atmosphere of Neptune: an Analysis of Radio Occultation Data Acquired with Voyager 2}",
      journal = {The Astronomical Journal},
         year = 1992,
        month = mar,
       volume = {103},
        pages = {967},
          doi = {10.1086/116119},
       adsurl = {https://ui.adsabs.harvard.edu/abs/1992AJ....103..967L}
}

@article{Hanel1981,
author = {Hanel, R. A. and Conrath, B. J. and Herath, L. W. and Kunde, V. G. and Pirraglia, J. A.},
title = {Albedo, internal heat, and energy balance of Jupiter: Preliminary results of the Voyager Infrared Investigation},
journal = {Journal of Geophysical Research: Space Physics},
volume = {86},
number = {A10},
pages = {8705-8712},
doi = {https://doi.org/10.1029/JA086iA10p08705},
url = {https://agupubs.onlinelibrary.wiley.com/doi/abs/10.1029/JA086iA10p08705},
eprint = {https://agupubs.onlinelibrary.wiley.com/doi/pdf/10.1029/JA086iA10p08705},
year = {1981}
}

@article{Militzer_2025,
doi = {10.3847/1538-4357/adef48},
url = {https://doi.org/10.3847/1538-4357/adef48},
year = {2025},
month = {aug},
publisher = {The American Astronomical Society},
volume = {990},
number = {1},
pages = {20},
author = {Militzer, Burkhard},
title = {Ab Initio Entropy Calculations of Water Predict the Interiors of Uranus and Neptune to Be 15%–30% Colder than Previous Models},
journal = {The Astrophysical Journal},
}

@INPROCEEDINGS{hubbard1995,
       author = {{Hubbard}, W.~B. and {Podolak}, M. and {Stevenson}, D.~J.},
        title = "{The interior of Neptune.}",
    booktitle = {Neptune and Triton},
         year = 1995,
       editor = {{Cruikshank}, Dale P. and {Matthews}, Mildred Shapley and {Schumann}, A.~M.},
        month = jan,
        pages = {109-138},
       adsurl = {https://ui.adsabs.harvard.edu/abs/1995netr.conf..109H}
}

@INCOLLECTION{podolak1991,
       author = {{Podolak}, M. and {Hubbard}, W.~B. and {Stevenson}, D.~J.},
        title = "{Models of Uranus' interior and magnetic field.}",
    booktitle = {Uranus},
         year = 1991,
       editor = {{Bergstralh}, Jay T. and {Miner}, Ellis D. and {Matthews}, Mildred S.},
        pages = {29-61},
       adsurl = {https://ui.adsabs.harvard.edu/abs/1991uran.book...29P}
}

@article{wang2025,
author = {Wang, Xinyue and Li, Liming and Roman, Michael and Zhang, Xi and Jiang, Xun and Fry, Patrick and Li, Cheng and Milcareck, Gwenael and Sanchez-Lavega, Agustin and Perez-Hoyos, Santiago and Hueso, Ricardo and Guillot, Tristan and Nixon, Conor and Dyudina, Ulyana and West, Robert and Kenyon, Matthew},
title = {Internal Heat Flux and Energy Imbalance of Uranus},
journal = {Geophysical Research Letters},
volume = {52},
number = {14},
pages = {e2025GL115660},
doi = {https://doi.org/10.1029/2025GL115660},
url = {https://agupubs.onlinelibrary.wiley.com/doi/abs/10.1029/2025GL115660},
eprint = {https://agupubs.onlinelibrary.wiley.com/doi/pdf/10.1029/2025GL115660},
note = {e2025GL115660 2025GL115660},
year = {2025}
}

@article{tejada-arevalo2025,
doi = {10.3847/2041-8213/adf3a5},
url = {https://doi.org/10.3847/2041-8213/adf3a5},
year = {2025},
month = {aug},
publisher = {The American Astronomical Society},
volume = {989},
number = {2},
pages = {L40},
author = {Tejada Arevalo, Roberto},
title = {Different Inhomogeneous Evolutionary Histories for Uranus and Neptune},
journal = {The Astrophysical Journal Letters},
}

@article{stevenson2026,
author = {Stevenson, David J.},
title = {Is the Uranus Heat Flow Puzzling?},
journal = {Geophysical Research Letters},
volume = {53},
number = {7},
pages = {e2026GL121709},
doi = {https://doi.org/10.1029/2026GL121709},
url = {https://agupubs.onlinelibrary.wiley.com/doi/abs/10.1029/2026GL121709},
eprint = {https://agupubs.onlinelibrary.wiley.com/doi/pdf/10.1029/2026GL121709},
note = {e2026GL121709 2026GL121709},
year = {2026}
}

@ARTICLE{Mazevet2019,
       author = {{Mazevet}, S. and {Licari}, A. and {Chabrier}, G. and {Potekhin}, A.~Y.},
        title = "{Ab initio based equation of state of dense water for planetary and exoplanetary modeling}",
      journal = {\aap},
         year = 2019,
        month = jan,
       volume = {621},
          eid = {A128},
        pages = {A128},
          doi = {10.1051/0004-6361/201833963},
archivePrefix = {arXiv},
       eprint = {1810.05658},
 primaryClass = {astro-ph.EP},
       adsurl = {https://ui.adsabs.harvard.edu/abs/2019A&A...621A.128M}
}

@article{French2016,
  title = {Ab initio calculation of thermodynamic potentials and entropies for superionic water},
  author = {French, Martin and Desjarlais, Michael P. and Redmer, Ronald},
  journal = {Phys. Rev. E},
  volume = {93},
  issue = {2},
  pages = {022140},
  numpages = {11},
  year = {2016},
  month = {Feb},
  publisher = {American Physical Society},
  doi = {10.1103/PhysRevE.93.022140},
  url = {https://link.aps.org/doi/10.1103/PhysRevE.93.022140}
}

@ARTICLE{Li2018,
       author = {{Li}, Liming and {Jiang}, X. and {West}, R.~A. and {Gierasch}, P.~J. and {Perez-Hoyos}, S. and {Sanchez-Lavega}, A. and {Fletcher}, L.~N. and {Fortney}, J.~J. and {Knowles}, B. and {Porco}, C.~C. and {Baines}, K.~H. and {Fry}, P.~M. and {Mallama}, A. and {Achterberg}, R.~K. and {Simon}, A.~A. and {Nixon}, C.~A. and {Orton}, G.~S. and {Dyudina}, U.~A. and {Ewald}, S.~P. and {Schmude}, R.~W.},
        title = "{Less absorbed solar energy and more internal heat for Jupiter}",
      journal = {Nature Communications},
         year = 2018,
        month = sep,
       volume = {9},
          eid = {3709},
        pages = {3709},
          doi = {10.1038/s41467-018-06107-2},
       adsurl = {https://ui.adsabs.harvard.edu/abs/2018NatCo...9.3709L}
}

@article{TOMASKO1978,
title = {Photometry and polarimetry of Jupiter at large phase angles: I. Analysis of imaging data of a prominent belt and a zone from pioneer 10},
journal = {Icarus},
volume = {33},
number = {3},
pages = {558-592},
year = {1978},
issn = {0019-1035},
doi = {https://doi.org/10.1016/0019-1035(78)90191-4},
url = {https://www.sciencedirect.com/science/article/pii/0019103578901914},
author = {M.G. Tomasko and R.A. West and N.D. Castillo},
}

@ARTICLE{irwin2025,
       author = {{Irwin}, Patrick G.~J. and {Wenkert}, Daniel D. and {Simon}, Amy A. and {Dahl}, Emma and {Hammel}, Heidi B.},
        title = "{The bolometric Bond albedo and energy balance of Uranus}",
      journal = {\mnras},
         year = 2025,
        month = jun,
       volume = {540},
       number = {2},
        pages = {1719-1729},
          doi = {10.1093/mnras/staf800},
archivePrefix = {arXiv},
       eprint = {2502.18971},
 primaryClass = {astro-ph.EP},
       adsurl = {https://ui.adsabs.harvard.edu/abs/2025MNRAS.540.1719I}
}

@ARTICLE{helled2010,
       author = {{Helled}, Ravit and {Anderson}, John D. and {Schubert}, Gerald},
        title = "{Uranus and Neptune: Shape and rotation}",
      journal = {Icarus},
         year = 2010,
        month = nov,
       volume = {210},
       number = {1},
        pages = {446-454},
          doi = {10.1016/j.icarus.2010.06.037},
archivePrefix = {arXiv},
       eprint = {1006.3840},
 primaryClass = {astro-ph.EP},
       adsurl = {https://ui.adsabs.harvard.edu/abs/2010Icar..210..446H}
}

@ARTICLE{nett2012,
       author = {{Nettelmann}, N. and {Becker}, A. and {Holst}, B. and {Redmer}, R.},
        title = "{Jupiter Models with Improved Ab Initio Hydrogen Equation of State (H-REOS.2)}",
      journal = {ApJ},
         year = 2012,
        month = may,
       volume = {750},
       number = {1},
          eid = {52},
        pages = {52},
          doi = {10.1088/0004-637X/750/1/52},
archivePrefix = {arXiv},
       eprint = {1109.5644},
 primaryClass = {astro-ph.EP},
       adsurl = {https://ui.adsabs.harvard.edu/abs/2012ApJ...750...52N}
}

@article{Kurosaki17,
   title={Acceleration of Cooling of Ice Giants by Condensation in Early Atmospheres},
   volume={153},
   ISSN={1538-3881},
   url={http://dx.doi.org/10.3847/1538-3881/aa6faf},
   DOI={10.3847/1538-3881/aa6faf},
   number={6},
   journal={The Astronomical Journal},
   publisher={American Astronomical Society},
   author={Kurosaki, Kenji and Ikoma, Masahiro},
   year={2017},
   month=may, pages={260} }

@article{Atreya20,
  author = {Atreya, S.K. and Hofstadter, M.H. and In, J.H. and Mousis, O. and Reh, K. and Wong, M.H.},
    Journal = {SSRv},
    Year = 2020,
    volume = {216},  pages = {18},
    Title = "{Deep Atmosphere Composition, Structure, Origin, and Exploration, with Particular Focus  on Critical in situ Science at the Icy Giants}",
    doi = {10.1007/s11214-020-0640-8}
}

@article{HMacF80, 
         Author={Hubbard, W. B. and MacFarlane, J. J.}, 
         Year={1980}, 
         Journal={J. Geophys. Res.}, 
         Volume={88}, Pages={225},
         Title="{Structure and Evolution of Uranus and Neptune}"}

@ARTICLE{canoamoros2024,
       author = {{Cano Amoros}, M. and {Nettelmann}, N. and {Tosi}, N. and {Baumeister}, P. and {Rauer}, H.},
        title = "{H$_{2}$─H$_{2}$O demixing in Uranus and Neptune: Adiabatic structure models}",
      journal = {\aap},
         year = 2024,
        month = dec,
       volume = {692},
          eid = {A152},
        pages = {A152},
          doi = {10.1051/0004-6361/202452148},
archivePrefix = {arXiv},
       eprint = {2410.21099},
 primaryClass = {astro-ph.EP},
       adsurl = {https://ui.adsabs.harvard.edu/abs/2024A&A...692A.152C}
}

@article{PEARL1990,
title = {The albedo, effective temperature, and energy balance of Uranus, as determined from Voyager IRIS data},
journal = {Icarus},
volume = {84},
number = {1},
pages = {12-28},
year = {1990},
issn = {0019-1035},
doi = {https://doi.org/10.1016/0019-1035(90)90155-3},
url = {https://www.sciencedirect.com/science/article/pii/0019103590901553},
author = {J.C. Pearl and B.J. Conrath and R.A. Hanel and J.A. Pirraglia and A. Coustenis},
}

@INCOLLECTION{guillot2015,
       author = {{Guillot}, T. and {Gautier}, D.},
        title = "{Giant Planets}",
    booktitle = {Treatise on Geophysics},
         year = 2015,
       editor = {{Schubert}, Gerald},
        pages = {529-557},
          doi = {10.1016/B978-0-444-53802-4.00176-7},
       adsurl = {https://ui.adsabs.harvard.edu/abs/2015trge.book..529G}
}

@article{Jacobson_2025,
doi = {10.3847/1538-3881/ad99d1},
url = {https://doi.org/10.3847/1538-3881/ad99d1},
year = {2025},
month = {jan},
publisher = {The American Astronomical Society},
volume = {169},
number = {2},
pages = {65},
author = {Jacobson, Robert A. and Park, Ryan S.},
title = {The Orbits of Uranus, Its Satellites and Rings, the Gravity Field of the Uranian System, and the Orientation of the Poles of Uranus and Its Satellites*},
journal = {The Astronomical Journal},
}

@article{Fortney_2011,
doi = {10.1088/0004-637X/729/1/32},
url = {https://doi.org/10.1088/0004-637X/729/1/32},
year = {2011},
month = {feb},
publisher = {The American Astronomical Society},
volume = {729},
number = {1},
pages = {32},
author = {Fortney, J. J. and Ikoma, M. and Nettelmann, N. and Guillot, T. and Marley, M. S.},
title = {SELF-CONSISTENT MODEL ATMOSPHERES AND THE COOLING OF THE SOLAR SYSTEM'S GIANT PLANETS},
journal = {The Astrophysical Journal},
}

@ARTICLE{Hub1977,
       author = {{Hubbard}, W.~B.},
        title = "{The Jovian Surface Condition and Cooling Rate}",
      journal = {\icarus},
         year = 1977,
        month = feb,
       volume = {30},
       number = {2},
        pages = {305-310},
          doi = {10.1016/0019-1035(77)90164-6},
       adsurl = {https://ui.adsabs.harvard.edu/abs/1977Icar...30..305H}
}

@ARTICLE{Kraus17,
       author = {{Kraus}, D. and {Vorberger}, J. and {Pak}, A. and {Hartley}, N.J. and {Fletcher}, L.B. and {Frydrych}, S. 
		and {Galtier}, E. and {Gamboa}, E.J. and {Gericke}, D.O. and {Glenzer}, S.H. and {Granados}, E. and {MacDonald}, 
		M.J. and {MacKinnon}, A.J. and {McBride}, E.E. and {Nam}, I. and {Neumayer}, P. and {Roth}, M. and {Saunders}, A.M. 
		and {Schuster}, A.K. and {Sun}, P. and {van Driel}, T. and {D{\"o}ppner}, T. and {Falcone}, R.W.},
        title = "{Formation of diamonds in laser-compressed hydrocarbons at planetary interior conditions}",
journal = {Nature Astronomy},
year = 2017,
volume = {1}, pages = {606-611},
doi = {10.1038/s41550-017-0219-9}
}

@ARTICLE{canoamoros26,
       author = {{Cano Amoros}, M. and {Nettelmann}, N. and {Tosi}, N. and {Mazevet}, S.},
        title = "{Revised entropy of the AQUA equation of state}",
      journal = {\aap},
         year = 2026,
        month = apr,
       volume = {708},
          eid = {A324},
        pages = {A324},
          doi = {10.1051/0004-6361/202557580},
       adsurl = {https://ui.adsabs.harvard.edu/abs/2026A&A...708A.324C}
}

@book{Fliessbach,
         Author={Flie{\ss}bach,T.},
         Year={1995},
         Publisher={Spectrum Akademischer Verlag Heidelberg, Berlin, Oxford, 2nd Edition},
         Title="{Statistische Phyik, Lehrbuch der Theoretischen Physik IV}"
}

@article{Hirai2009,
         Author = {Hirai, Hisako and Konagai, Keisuke and Kawamura, Taro 
                 and Yamamoto, Yoshitaka and Yagi, Takehiko},
         Year = {2009},
         Journal = {Physics of the Earth and Planetary Interiors},
         Volume = {174}, pages = {242 - 246},
         Title = "{Polymerization and diamond formation from melting methane and 
                 their implications in ice layer of giant planets}",
         Doi = {10.1016/j.pepi.2008.06.011}
}

@ARTICLE{Frost2024,
       author = {Frost, M. and McWilliams, R.S. and Bykova, E. and {et al}},
        title = "{Diamond precipitation dynamics from hydrocarbons at icy planet interior conditions}",
      journal = {Nat. Ast.},
         year = 2024, volume = {8}, pages = {174},
          doi = {10.1038/s41550-023-02147-x}
}

@article{Bethkenhagen17,
         Author = {Bethkenhagen, M. and Meyer, E.R. and Hamel, S. and Nettelmann, N. and French, M. and Scheibe, L. 
         and Ticknor, C. and Coolins, L.A. and Kress, J.D. and Fortney, J.J. and Redmer, R.},
         Year = {2017},
         Journal = {ApJ}, Volume={848}, Pages={67},
         Title = "{Planetary Ices and the Linear Mixing Approximation}",
         doi = {10.3847/1538-4357/aa8b14}
}

@article{Holst08, 
         Author={Holst, Bastian and Redmer, Redmer and Desjarlais, M.P.}, 
         Title = {Thermophysical properties of warm dense hydrogen using quantum molecular dynamics simulations},
         Year={2008}, Journal={PRB}, 
         Volume={77}, Pages={184201},
         doi = {10.1103/PhysRevB.77.184201}}

@ARTICLE{Benedetti99,
       author = {{Benedetti}, L.~R. and {Nguyen}, J.~H. and {Caldwell}, W.~A. and {Liu}, H. and {Kruger}, 
			M. and {Jeanloz}, R.},
        title = "{Dissociation of CH$_{4}$ at high pressures and temperatures: diamond formation in giant planet interiors?}",
      journal = {Science},
         year = 1999,
       volume = {286}, pages = {100-102},
          doi = {10.1126/science.286.5437.100}
}

@ARTICLE{Preising23,
       author = {{Preising}, M. and {French}, M. and {Mankovich}, C. and {Soubiran}, F. and {Redmer}, R.},
        title = "{Material Properties of Saturn's Interior from Ab Initio Simulations}",
      journal = {ApJS},
         year = 2023,
       volume = {269},
        pages = {47},
          doi = {10.3847/1538-4365/ad0293}
}

@ARTICLE{Ranieri24,
       author = {Ranieri, U. and Formisano, F. and  Gorelli, F.A. and },
        title = "{Crossover from gas-like to liquid-like molecular diffusion in a simple supercritical fluid.}",
      journal = {Nat.~Comm.},
         year = 2024,
        pages = {4142},
          doi = {10.1038/s41467-024-47961-7}}

@article{Nettelmann08, 
         Author={Nettelmann, N. and Holst, B. and Kietzmann, A. and French, M. 
         and Redmer, R. and Blaschke, D.}, 
         Year={2008}, 
         Journal={Apj}, 
         Volume={683}, Pages={1217}}

@ARTICLE{Darafeyeu24,
       author = {{Darafeyeu}, Valiantsin and {Rimle}, Stephanie and {Mazzola}, Guglielmo and {Helled}, Ravit},
        title = "{The Linear Mixing Approximation in Silica─Water Mixtures at Planetary Conditions}",
      journal = {ApJ},
         year = 2024,
       volume = {975},
        pages = {255},
          doi = {10.3847/1538-4357/ad7e29}
}

@ARTICLE{Gupta25,
       author = {{Gupta}, Akash and {Stixrude}, Lars and {Schlichting}, Hilke E.},
        title = "{The Miscibility of Hydrogen and Water in Planetary Atmospheres and Interiors}",
      journal = {ApJL},
         year = 2025,
       volume = {982},
        pages = {L35},
          doi = {10.3847/2041-8213/adb631}
}

@ARTICLE{Vorberger07,
         author = {J.~Vorberger and I.~Tamblyn and B.~Militzer and S.~A.~Bonev},
         journal = {Phys. Rev. B},
         year = {2007},
         volume = {75}, pages = {024206},
         title = {Hydrogen-helium mixtures in the interiors of giant planets}}

@ARTICLE{Kovacevic23,
       author = {{Kova{\v{c}}evi{\'c}}, Tanja and {Gonz{\'a}lez-Cataldo}, Felipe and {Militzer}, Burkhard},
        title = "{The homogeneous mixing of MgO and H2O at extreme conditions}",
      journal = {Contributions to Plasma Physics},
         year = 2023,
       volume = {63}, number = {9-10}, pages = {e202300017},
          doi = {10.1002/ctpp.202300017}
}

@article{Militzer24,
         Author = {Militzer, B. },
         Year = {2024}, 
         Journal = {PNAS},
         Volume = {121}, Pages = {e2403981121},
         Title = "{Phase separation of planetary ices explains nondipolar magnetic
				fields of Uranus and Neptune.}",
         doi = {10.1073/pnas.2403981121} }

@ARTICLE{Militzer25,
       author = {{Militzer}, Burkhard},
        title = "{Ab Initio Entropy Calculations of Water Predict the Interiors of Uranus and Neptune to Be 15\%─30\% Colder than Previous Models}",
      journal = {ApJ},
         year = 2025, volume = {990}, number = {1},
        pages = {20},
          doi = {10.3847/1538-4357/adef48},
}

@ARTICLE{Nisr20,
       author = {{Nisr}, Carole and {Chen}, Huawei and {Leinenweber}, Kurt and {Chizmeshya}, 
		Andrew and {Prakapenka}, Vitali B. and {Prescher}, Clemens and {Tkachev}, 
			Sergey N. and {Meng}, Yue and {Liu}, Zhenxian and {Shim}, Sang-Heon},
        title = "{Large H$_{2}$O solubility in dense silica and its implications for the interiors 
		of water-rich planets}",
      journal = {Proceedings of the National Academy of Science},
         year = 2020,
       volume = {117},
       number = {18}, pages = {9747-9754},
          doi = {10.1073/pnas.1917448117}
}

@ARTICLE{Kim21,
       author = {{Kim}, Taehyun and {Chariton}, Stella and {Prakapenka}, Vitali and {Pakhomova}, 
		Anna and {Liermann}, Hanns-Peter and {Liu}, Zhenxian and {Speziale}, Sergio and {Shim}, 
		Sang-Heon and {Lee}, Yongjae},
        title = "{Atomic-scale mixing between MgO and H$_{2}$O in the deep interiors of water-rich planets}",
      journal = {Nature Astronomy},
         year = 2021, volume = {5}, pages = {815-821},
          doi = {10.1038/s41550-021-01368-2}
}

@ARTICLE{Stalder01,
       author = {{Stalder}, R. and {Ulmer}, P. and {Thompson}, A.~B. and {G{\"u}nther}, D.},
        title = "{High pressure fluids in the system MgO-SiO$_{2}$-H$_{2}$O under upper mantle conditions}",
      journal = {Contributions to Mineralogy and Petrology},
         year = 2001,
       volume = {140}, pages = {607-618},
          doi = {10.1007/s004100000212}
}

@ARTICLE{AGU2025,
       author = {{Mauritsen}, Thorsten and {Tsushima}, Yoko and {Meyssignac}, Benoit and {Loeb}, Norman G. and {Hakuba}, Maria and {Pilewskie}, Peter and {Cole}, Jason and {Suzuki}, Kentaroh and {Ackerman}, Thomas P. and {Allan}, Richard P. and {Andrews}, Timothy and {Bender}, Frida A.-M. and {Bloch-Johnson}, Jonah and {Bodas-Salcedo}, Alejandro and {Brookshaw}, Anca and {Ceppi}, Paulo and {Clerbaux}, Nicolas and {Dessler}, Andrew E. and {Donohoe}, Aaron and {Dufresne}, Jean-Louis and {Eyring}, Veronika and {Findell}, Kirsten L. and {Gettelman}, Andrew and {Gristey}, Jake J. and {Hawkins}, Ed and {Heimbach}, Patrick and {Hewitt}, Helene T. and {Jeevanjee}, Nadir and {Jones}, Colin and {Kang}, Sarah M. and {Kato}, Seiji and {Kay}, Jennifer E. and {Klein}, Stephen A. and {Knutti}, Reto and {Kramer}, Ryan and {Lee}, June-Yi and {McCoy}, Daniel T. and {Medeiros}, Brian and {Megner}, Linda and {Modak}, Angshuman and {Ogura}, Tomoo and {Palmer}, Matthew D. and {Paynter}, David and {Quaas}, Johannes and {Ramanathan}, Veerabhadran and {Ringer}, Mark and {Schuckmann}, Karina and {Sherwood}, Steven and {Stevens}, Bjorn and {Tan}, Ivy and {Tselioudis}, George and {Sutton}, Rowan and {Voigt}, Aiko and {Watanabe}, Masahiro and {Webb}, Mark J. and {Wild}, Martin and {Zelinka}, Mark D.},
        title = "{Earth's Energy Imbalance More Than Doubled in Recent Decades}",
      journal = {AGU Advances},
         year = 2025,
        month = jun,
       volume = {6},
       number = {3},
        pages = {e2024AV001636},
          doi = {10.1029/2024AV001636},
}

@book{FN09, 
         Author={Fortney, Jonathan James and Nettelmann, Nadine}, 
         Year={2010}, 
         Booktitle={Planetary Magnetism},
         Editor={Christensen, U.R. and Balogh, A. and Breuer, D. and Gla{\ss}meier, K.-H.},
         publisher={Springer Space Science Reviews}, 
         Volume={152}, Pages={423-447},
         Title={The Interior Structure, Composition, and Evolution of Giant Planets}}

@techreport{SESAME, 
         Author={Lyon, S.P. and Johnson, J. D. (eds.}, 
         Year={1992}, 
         Title={SESAME: Los Alamos National Laboratory Equation of State Database},
 	       Institution={LANL report no. LA-UR-92-3407}}

@article{Altshuler1965,
  author    = {Al'tshuler, L. V.},
  title     = {Application of shock waves in high pressure physics},
  journal   = {Uspekhi Fizicheskikh Nauk},
  volume    = {85},
  number    = {2},
  pages     = {197--258},
  year      = {1965},
  note      = {(in Russian)},
  language  = {russian}
}

@ARTICLE{Walzer1982,
       author = {{Walzer}, U.},
        title = "{A new equation of state for high compression}",
      journal = {Physics of the Earth and Planetary Interiors},
         year = 1982,
       volume = {30},     number = {1}, pages = {62-83},
          doi = {10.1016/0031-9201(82)90128-5}
}

@ARTICLE{He22,
       author = {{He}, Zhiyu and {R{\"o}del}, Melanie and {L{\"u}tgert}, Julian and {Bergermann}, Armin and {Bethkenhagen}, Mandy and {Chekrygina}, Deniza and {Cowan}, Thomas E. and {Descamps}, Adrien and {French}, Martin and {Galtier}, Eric and {Gleason}, Arianna E. and {Glenn}, Griffin D. and {Glenzer}, Siegfried H. and {Inubushi}, Yuichi and {Hartley}, Nicholas J. and {Hernandez}, Jean-Alexis and {Heuser}, Benjamin and {Humphries}, Oliver S. and {Kamimura}, Nobuki and {Katagiri}, Kento and {Khaghani}, Dimitri and {Lee}, Hae Ja and {McBride}, Emma E. and {Miyanishi}, Kohei and {Nagler}, Bob and {Ofori-Okai}, Benjamin and {Ozaki}, Norimasa and {Pandolfi}, Silvia and {Qu}, Chongbing and {Ranjan}, Divyanshu and {Redmer}, Ronald and {Schoenwaelder}, Christopher and {Schuster}, Anja K. and {Stevenson}, Michael G. and {Sueda}, Keiichi and {Togashi}, Tadashi and {Vinci}, Tommaso and {Voigt}, Katja and {Vorberger}, Jan and {Yabashi}, Makina and {Yabuuchi}, Toshinori and {Zinta}, Lisa M.~V. and {Ravasio}, Alessandra and {Kraus}, Dominik},
        title = "{Diamond formation kinetics in shock-compressed C─H─O samples recorded by 
		small-angle x-ray scattering and x-ray diffraction}",
      journal = {Science Advances},
         year = 2022, volume = {8}, number = {35},
        pages = {eabo0617},
          doi = {10.1126/sciadv.abo0617}
}

@article{LHR09,
         Author={Lorenzen, Winfried and Holst, Bastian and Redmer, Ronald},
         Year={2009},
         Journal={Phys. Rev. Lett.},
         Volume={102}, Pages={5701},
         Title={Demixing of Hydrogen and Helium at Megabar Pressures}
}

\begin{appendix}

\section{Adjusted CH$_4$ and He abundances}\label{apx:meth_eqs}

This section describes how we determine the mass fractions of methane, helium, and water in the full mixture given a $\rm H_2O/H_2$ ratio that has been determined from the H$_2$/H$_2$O phase diagram, see \cite{canoamoros2024}. 
We assume four given parameters. Those are $\Tone$, the helium mass fraction $Y^{(2)}$ in the two-component H-He system, which is assumed to remain at a protosolar value of 0.275, the methane mass fraction $\Zmeth$ in the current atmosphere, and the water particle fraction $x_{\rm H2O}^{(2)}$ in the water-hydrogen system.
The latter parameter is obtained from the water-hydrogen phase diagram, where it is defined as
$x_{\rm H2O}^{(2)} = N_{\rm H2O}\,/\,(N_{\rm H2O} + N_{\rm H2})$. This definition looks as if it was implicitly assumed that all hydrogen occurs in molecular form in either H$_2$O or H$_2$. Now in the hydrogen-helium phase diagram, e.g., \citep{LHR09}, the helium particle fraction is defined as $x_{\rm He}^{(2)} = N_{\rm He}\,/\,(N_{\rm He} + N_{\rm H})$. That definition looks as if it was implicitly assumed that all hydrogen occurs in atomic form. These definitions are  inherited from the MD-simulations that are employed to compute the phase diagrams. There, the total number of hydrogen, helium, and oxygen particles is counted regardless in which molecular form they occur.  Here we want to refer to chemical species so that particle ratios remain finite when all hydrogen is in molecular or atomic form. Thus we write 
\begin{equation}
    x_{\rm H2O}^{(2)} = x_{\rm H2O}^{(sep)} = N_{\rm H2O}\,/\,(N_{\rm H2O} + N_{\rm H2} + 0.5\,N_{\rm H})
\end{equation}
where particle numbers $N_i$ refer to chemical species. We consider the species H, H$_2$, He, H$_2$O, and CH$_4$. 
With their molar masses $m_i$  we obtain  the water mass fraction in the water-hydrogen system
\begin{equation}\label{eq.xZH2O}
    \Zwaterzwei(\Tone) = \frac{m_{\rm H2O}\,\xwaterzwei}{m_{\rm H2O}\,\xwaterzwei + \mHzwei\,\left(1-\xwaterzwei\right)}
\end{equation}

The wanted parameters are the methane and helium mass fractions $\Zmeth$ and $Y$ when $\xwaterzwei$ changes with $\Tone$.
We assume that the H$_2$/H, He/H and CH$_4$/H$_2$-ratios remain constant over time in the outer envelope while the H$_2$O/H$_2$ ratio changes due to the water-hydrogen phase separation at depth.
We set the present CH$_4$/H$_2$ ratio to be consistent with the observed atmospheric methane abundance of 50--90$\times$ solar in both planets atmospheres \citep{Atreya20}. 
Within that range, we choose a value $\Zmeth\sim$0.1 that gives a good agreement with the observed gravitational harmonics $J_2$, $J_4$, and the equatorial radius for the present state with $\Zwaterzwei$.

We furthermore introduce $\NHstar \:=\: \NH + \NHzwei$ and the degree of hydrogen dissociation 
$\alpha_{\rm H} = \NH/\NHstar$, from which it follows 
\begin{equation}\label{eq:NNN_HHH}
    \NH/\NHstar = \alpha\quad,\quad \NHzwei/\NHstar = 1-\alpha_H\:,
\end{equation}
and $\NH\:(1-\alpha_H) = \NHzwei\:\alpha_H\:$.
The quantities we are aiming at are 
\begin{equation}\label{eq:ZmethZwat}
	\Zmeth(\Tone) = \frac{N_{\rm CH4}/\NHstar}{ \Sigma_i / m_{\rm CH4} }\quad\mbox{(a)}\:,\quad
	\Zwater(\Tone) = \frac{N_{\rm H2O}/\NHstar}{ \Sigma_i / m_{\rm H2O} }\quad\mbox{(b)}
\end{equation}
with 
$\Sigma_i = m_i \: (N_i\NHstar)$ and $i=\{\rm H,H_2,He,H_2O,CH_4\}$. 
We determine 
\begin{equation}\label{eq:NNN_H2O}
	\frac{ N_{\rm H2O} }{\NHstar} = \frac{\Zwaterzwei(\Tone)}{1-\Zwaterzwei(\Tone)} \: 
		\left( \frac{\mH}{m_{\rm H2O}}\,\alpha_H + \frac{\mHzwei}{m_{\rm H2O}} (1-\alpha_H) \right)
\end{equation}
\begin{equation}\label{eq:NNN_He}
	\frac{\NHe}{\NHstar} = \frac{Y^{(2)}}{1-Y^{(2)}} \: 
		\left(\frac{\mH}{\mHe}\,\alpha_H + \frac{\mHzwei}{\mHe} (1-\alpha_H)\right)\:=\:\mbox{const.}_t
\end{equation}
\begin{eqnarray}\label{eq:NNN_CH4}
	\frac{N_{\rm CH4}}{\NHstar} &=& \frac{\Zmeth}{1-\Zmeth} \: 
	\Bigg[ \frac{\mH}{m_{\rm CH4}}\,\alpha_H + \frac{\mHzwei}{m_{\rm CH4}}\,(1-\alpha_H)
\nonumber\\
		&+& 
		\left( 	\frac{1}{m_{\rm CH4}}\, \frac{Y^{(2)}}{1-Y^{(2)}} 
				+ \frac{1}{m_{\rm CH4}}\, \frac{\Zwaterzwei}{1-\Zwaterzwei} \right)
\nonumber\\
		&& \times  \:\left(\: \mH\,\alpha_H + \mHzwei (1-\alpha_H) \:\right)\:  
	\Bigg] \:=\:\mbox{const.}_t
\end{eqnarray}
Inserting expressions (\ref{eq:NNN_HHH}, \ref{eq:NNN_H2O}, \ref{eq:NNN_He}, \ref{eq:NNN_CH4})
into (\ref{eq:ZmethZwat}a,b) yields computable expressions.
Finally, with $Z_1(\Tone) = \Zwater(\Tone) + \Zmeth(\Tone)$ the helium and hydrogen mass fractions in the full mixture can be obtained as $Y_1(t) = Y^{(2)} \: (1-Z_1(t))$ and $X_1=1-Y_1-Z_1$, respectively.

In the computation of the abundances, we set $\alpha_H=0$. We estimate $\alpha_H \lesssim 0.2$ since the temperatures in the outer envelope must not exceed 2000 K for H$_2$/H$_2$O phase separation to occur and pressures there do not exceed 11 GPa. Under these conditions, $\rho_{H} \lesssim 0.28\:$g/cc, for which \citet{Holst08} find  $\NHzwei/\left(0.5 \NH + \NHzwei\right) > 0.8$ yielding $\alpha_H<0.33$ and decreasing with lower $T$ and lower $P$ as in the outer envelopes.    
\section{The Methane EoS CH4-REOS.1}\label{apx:methane}

\subsection{Construction}

The methane EoS includes DFT-MD data, Sesame-5550 data, and ideal gas data. We took the table of methane EoS data whose $P$-$\rho$-isotherms were illustrated in \citet{nettelmann16} and applied to the icy Uranus model in \citet{Bethkenhagen17} and added more data as well as the absolute entropy. That earlier table was prepared in 2014 and ranged from $T=$ 2000 to 10000$\:$K and $\rho=0.03$ (few GPa) to 4.0 g/cc ($\sim$10 Mbar). While the underlying DFT-MD data were provided to us by S.~Hamel (pers.comm.~2014) and are consistent to those described in \citet{Bethkenhagen17}, a description of the larger-scale methane EoS was never published. Here, we describe the construction of the updated large-scale methane EoS, hereafter CH4-REOS.1.

\emph{Sesame to DFT-MD}~
The coldest isotherm of the Sesame-5550 table is for 270 K; the warmest one that is at our disposal is for 10000 K while the density range is 0.105---2.5 g/cc. We calculate isotherms by interpolation on the Sesame-5550 EoS density grid for temperatures of 500, 1000, 1500, 2000, 2500, 3000, 4000, $\ldots$, to 10000~K. Only these Sesame-EOS based isotherms enter CH4-REOS. The isotherms for $1000\leq T\leq 10000$ K were connected to the available DFT-MD data at these temperatures by interpolation in $P(\rho)$ and $u(\rho)$. 

\emph{High densities}~
The highest densities that is provided by the DFT-MD data is 4.0 g/cc with maximum pressures of 10 Mbar. We added additional points at 6 and 12 g/cc by forcing all isotherms to converge to the point 100 Mbar at 12 g/cc, which was chosen to yield a smooth slope. A merging behavior of isotherms toward high densities  is known from other substances such as hydrogen and water, where it results from the increasing influence of the Fermi pressures from pressure-ionized electrons on the total pressure. The internal energies at the extrapolated points 6 and 12 g/cc were subsequently adjusted to yield the behavior of an ideal quantum gas, $u\sim P/\rho$ except for the low-T isotherms.

\emph{Low densities: Sesame to ideal gas}~
The isotherms for 500---10000 K were extended toward lower densities
by connecting them to the ideal gas EoS, where an ideal gas of methane particles without internal
degrees of freedom was assumed (i.e. $U = 3/2\, N\, k_B T)$. The $P$--$\rho$ relations of the ideal gas isotherms transition quite well to the Sesame-isotherms, although a little interpolation was necessary for a smooth transition. The interpolated region for the $P-\rho$ relation is 0.050--0.105 g/cc, and further 0.6--0.81 g/cc to remove a local maximum in the Sesame data along the 500-K isotherm, while 0.010---0.105 for all higher-$T$ isotherms.
For all isotherms, the internal energy of the ideal gas EoS was adjusted to the values of the Sesame EoS at 0.05 g/cc and then the transition be smoothed. 

\emph{Low and high temperatures}~
The warmest isotherm is for 20000 K. It is solely extrapolated from cooler isotherms along isochores.
At $60 \leq T< 500$~K, only the ideal gas EoS was used. Its internal energy levels were adjusted to maintain their offset to the 500 K isotherm. To avoid extremely high densities at high pressures, also these cold isotherms are forced to merge to the 100$\,$Mbar-at-12 g/cc point. Assuming an ideal gas at low-T and pressures of several hundred bars violates the known phase diagram. For instance, \citet{Ranieri24} find experimental evidence of liquid and solid phases at T < 100 K and P > 1 bar, which in the atmospheres of Uranus and Neptune lead to methane clouds at $\sim$1--1.4 bars \citep{Atreya20}. With a critical  point at 190 K and 46 bars and liquid and solid phases underneath, one may wonder why there we assumed the gas phase? The reason is that this improves the accuracy of the calculated density of the mixture if one uses Amagat's law of linear mixing ($\rho_i(P,T)$), as it is the common practice in the modeling of fluid planets. Amagat's law can significantly over-estimate the density of a substance if it is in the solid phase according to given total pressure but would be in the gas phase according to its partial pressure in a given total volume. This difference can clearly be seen in Figure \ref{fig:methane_PUrho} when comparing the 100 K water isotherm, which does account for the liquid and ice phases in pure water, to the methane-line at 100~K.
This methane EoS is recommended for  users who do not wish to switch from Amagat's law to Dalton's law under the clouds.

\begin{figure*}
\centering
\includegraphics[width=0.47\textwidth]{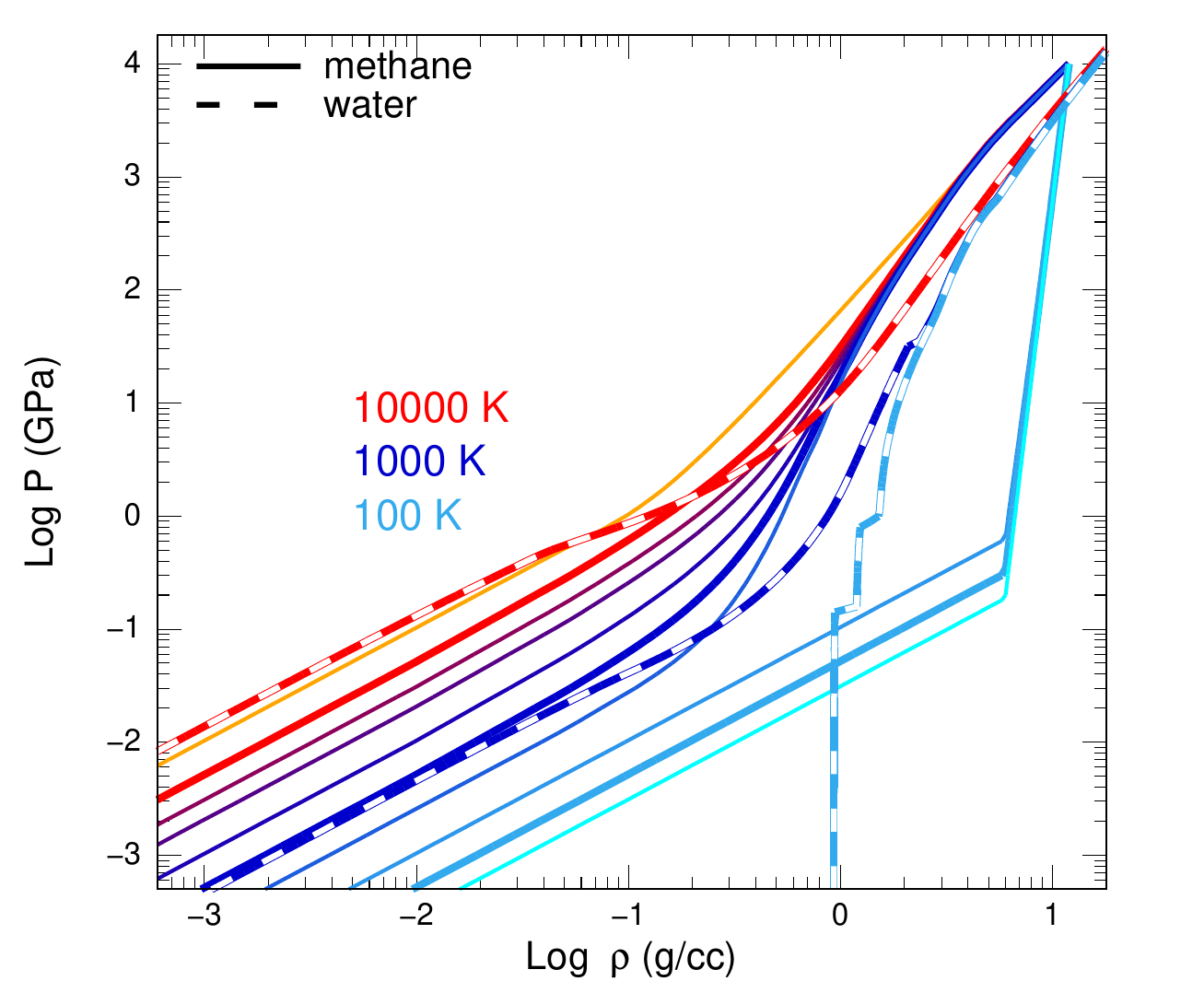}
\includegraphics[width=0.46\textwidth]{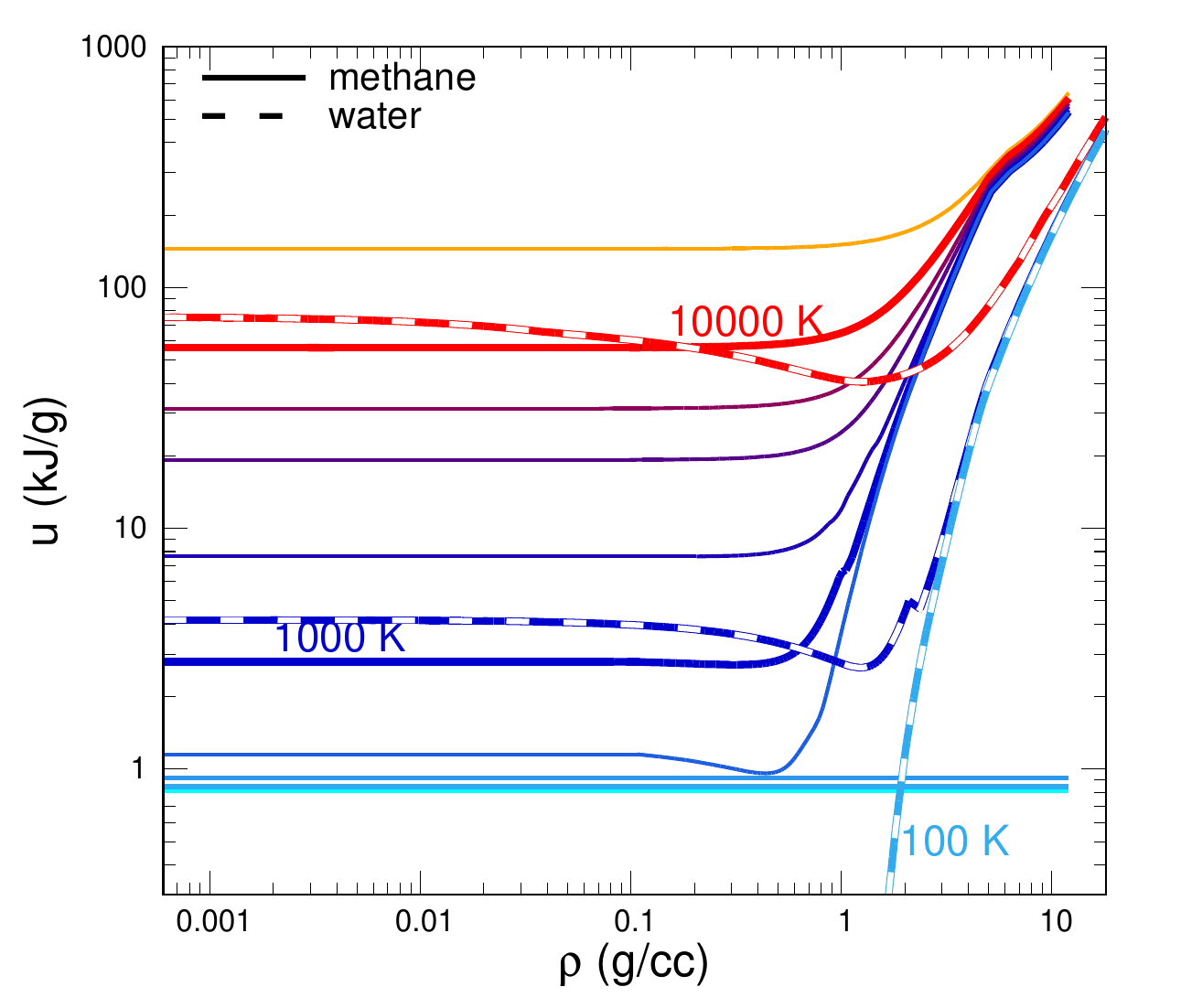}
\caption{Pressure (left) and Internal energy (right) over density along isotherms of the CH4-REOS.1 methane EoS (solid lines) for temperatures (in K) of 60, 100, 200, 500, 1k, 2k, 4k, 6k, 10k, 20k and of the Revised AQUA water EOS (dashed lines) for 100 K (cyan), 1000 K (blue), 10000 K (red).   }
\label{fig:methane_PUrho}
\end{figure*}

\emph{Entropy}~
Finally, the specific entropy $s+s_0$ was obtained by thermodynamic integration (TDI) using a reference point of 0.1 g/cc at 1000 K. The unknown offset $s_0$ was adjusted so that the absolute values $s$ reproduce the Sackur-Tetrode relation in the ideal gas limit, that is for isotherms 60---500 K.

\subsubsection{Absolute entropy from Sackur-Tetrode}

We gauge the entropy of the methane EoS to that of a monoatomic one-component ideal gas, in which case the absolute entropy is given by the Sackur-Tetrode relation
\begin{equation}\label{eq:sackurtetrode}
	s = \big(N_A\,k_B/\mmol\big)\:\left( 5/2  +  \ln\left(\frac{k_BT}{P}\lambdath^{-3}\right) \right)
\end{equation}
with the thermal wavelength
$
	\lambdath = \sqrt{ (2\pi\hbar^2)\,/\,(m_1 k_B T) }
$
where  $m_1$ is the mass of a single particle, $\mmol$ is the molar mass, $N_A$ is Avogadro's constant, $k_B$ is Boltzmann's constant, and $2\pi \hbar=h$ and $h$ is Planck's constant.
The derivation of Equation~(\ref{eq:sackurtetrode}) in the following relies on equations given in \cite{Fliessbach}, hereafter [FSP].
The absolute entropy of a system of volume $V$, pressure $P$, temperature $T$, and containing $N$ particles is given by the Gibbs-Duhem Equation,
\begin{equation}\label{eq:GibbsDuhem}
	S = \frac{U}{T} + \frac{PV}{T} - \frac{\mu N}{T}
\end{equation}
where $\mu$ is the chemical potential and $PV=N\,k_B\,T$ is the ideal gas equation of state. While entropy $S$ and internal energy $U$ in Eq.~(\ref{eq:GibbsDuhem}) scale with the system size, we are interested in the specific entropy per gram, 
\begin{equation}\label{eq:littles}
	s = S/N\times N_A/\mmol \:. 
\end{equation}
The task is to find analytical expressions for $\mu$ and $U$ so that Eq.~(\ref{eq:GibbsDuhem}) can be evaluated. 
The chemical potential $\mu$ is derived from the Helmholtz free energy $F(T,V,N) = -k_B T\,\ln Z$ [FSP]
where $Z(T,V,N)$ is the sum of states of the system. The differential for $F$ is (Gibbs' Fundamental Equation)
$dF = -S\,dT - P\,dV + \mu\,dN$, thus $\mu = dF/dN$. For an ideal gas, $Z$ can be factorized to $Z(T,V,N) = \frac{1}{N!}\: Z_1(T,V)^N$ \: where the pre-factor $1/N!$  reduces the number of states to only those that are distinct from each other. The 1-particle sum of states is given by [FSP],
\begin{equation}\label{eq:Z1}
	Z_1 = \frac{1}{(2\pi\hbar)^3}\int\!\! d^3r\!\! \int d^3p\: \exp\{-\beta H_1(\mathbf{r,p})\}
\end{equation}
where $H_1(\mathbf{r,p})$ is the Hamilton function for one particle. For a molecule that can move around freely, rotate and vibrate, the classical Hamilton function can be written (Eq.~27.2 in [FSP]) $H_1 = H_{\rm trans} + H_{\rm rot} + H_{\rm vib}\:$. For a single atom, $H_1$ reduces to the kinetic energy due to translational motion in three directions $H_{trans} = (p_x^2 + p_y^2 + p_z^2)/(2m_1)$.
The $f=$ three degrees of freedom contribute $k_BT/2$ each to the internal energy of the system, as 
would do any other degree of freedom on average. From $U=(f/2)Nk_B T$ follows 
\begin{equation}\label{eq:littleu}
	u=\frac{3}{2}\: \frac{N_A}{\mmol}\: k_BT\:.
\end{equation}
Inserting $H_{\rm trans}$ into (\ref{eq:Z1}) yields
\[
	Z_1 = 2\pi\:\frac{V\,(2m_1\, k_B T)^{3/2}}{(2\pi\hbar)^3} \int_0^{\infty} dt\,t^{1/2} e^{-t} \\
\]
with $t=p^2/(2m_1 k_B T)$. The integrand can be expressed in terms of the Gamma-function  
$
	\Gamma(r+1) = \int_0^{\infty} dt\: t^{r}\:e^{-t}
$ (Eq.~3.25 in [FSP]). For natural numbers $n$,
$
	\Gamma(1/2+n) = (\,(2n)!/(4^n n!)\,)\:\sqrt{\pi}	
$.
Here, $r=1/2$ and $n=1$, thus 
\begin{equation}\label{eq:Z1fin}
	Z_1 = \frac{V\,(2\pi m_1\, k_B T)^{3/2}}{(2\pi\hbar)^3} \:= \frac{V}{\lambdath^3}\:.
\end{equation}
Now for $N$ particles in $V$ and translations only,
$
	Z(T,V,N) = (1/N!)\: \left( V/\lambdath^3 \right)^N
$ (Eq.~25.7 in [FSP]).
In the limit of large N, $\ln N!\; = N \ln N - N$ (Stirling formula) and thus
$
	\ln Z(T,V,N) = N + N\ln\left( (V/N)\lambda^{-3}\right)
$. Its derivative is 
$	
 d\ln Z/dN = \ln\left( (V/N) \lambda^{-3}\right)
$.
Finally the chemical potential of an monoatomic ideal gas is obtained as 
\begin{equation}\label{eq:mutrans}
	\mu = -k_B T\: \ln\left(\frac{V}{N}\lambda^{-3}\right)\quad.
\end{equation}
Inserting Eqs.~(\ref{eq:littleu}) and (\ref{eq:mutrans}) into (
\ref{eq:GibbsDuhem}) and (\ref{eq:GibbsDuhem}) into (\ref{eq:littles}) yields (\ref{eq:sackurtetrode}).

\subsection{Behavior of Entropy $S(T,P)$ }

\begin{figure}
\centering
\includegraphics[width=0.5\textwidth]{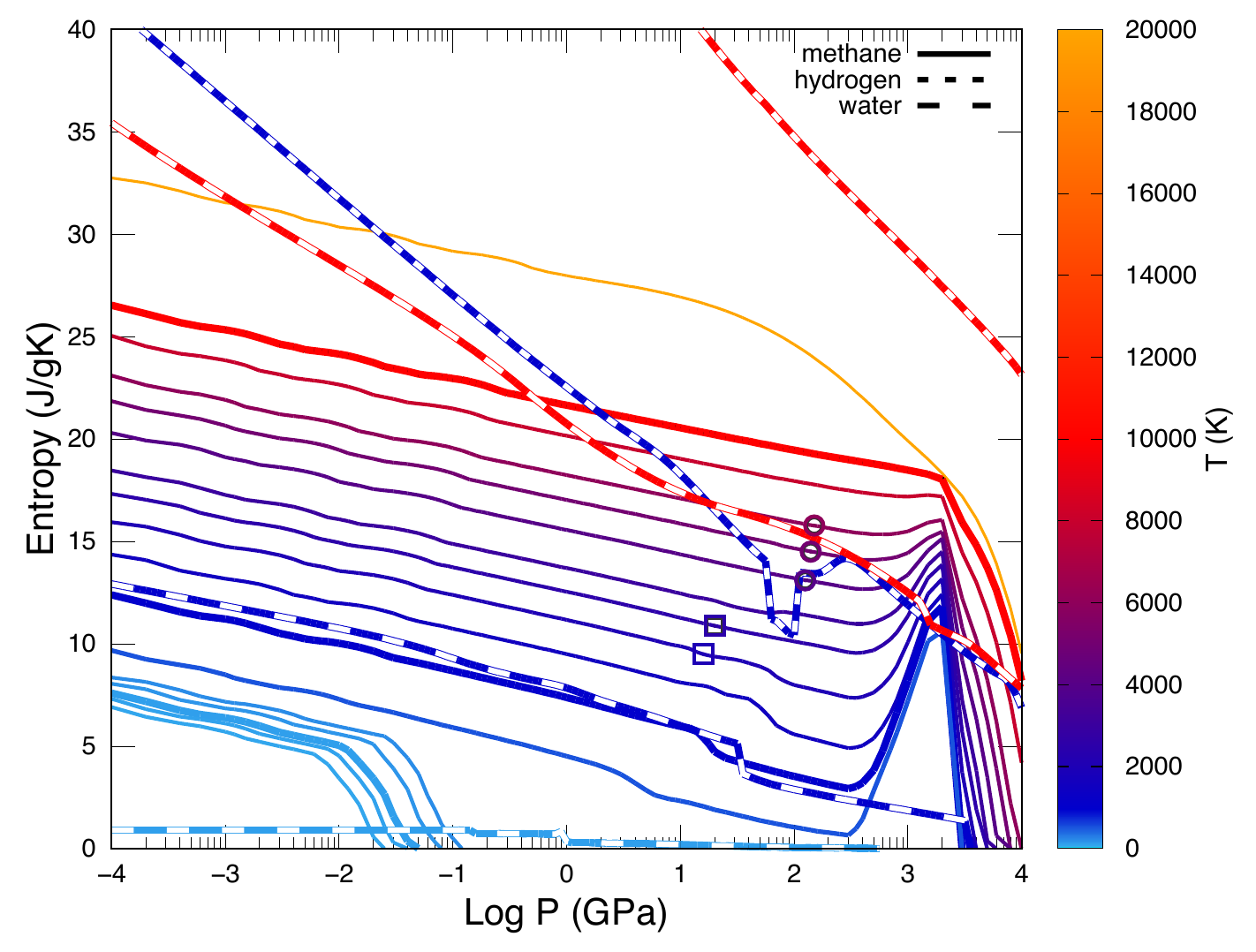}
\caption{Entropy along isotherms of the methane EoS (solid lines) for (botton-to-top) $T=$20, 10, 8, 6, 5, 4, 3, 2.5, 2, 1.5, 1 kK, 500, 200, 150, 100, 80, 60 K, of the CMS19 hydrogen EOS (short-dashed) for 10, 1, kK, and of the Revised AQUA water EOS (long-dashed) for 10, 1, 0.1 kK; symbols show the $P,T$ starting conditions for phase separation into diamond and hydrogen seen in static DAC experiments (squares) and dynamic shock compression experiments (circles). Symbols are placed at the $s(P,T)$ conditions of the methane EoS. Thick lines highlight isotherms for 100, 1000, 10000 K.}
\label{fig:methane_entropy}
\end{figure}

Figure \ref{fig:methane_entropy} shows the specific entropy $s(\rho)$ along isotherms.
There are six striking features; first, $ds/dT_{P} > 0$ along isobars, second $ds/dP_{T} < 0$ along isotherms except for third, a region at $P \sim$6---10 Mbar where $ds/dP_{|T} > 0$ for isotherms $500 \lesssim T \lesssim 8000$~K 
beyond which (fourth), starting at $\sim 20$ Mbar a steep decline occurs; further fifth, a shallow dip 
for isotherms $500 \lesssim T \lesssim 2000$~K in the 10--100 GPa region, and sixth, 
a strong dip to zero for cold isotherms $<500$ K.

Around feature four, we manually enforced otherwise overlapping isotherms to behave monotonously with $T$ in order 
to ensure that specific heat $c= T\;ds/dT$ remains  positive.

Feature two is expected. From the Maxwell-relation $dS/dP_T = -dV/dT_{P}$,
$dV/d\rho = -V^2/m$, and $s = S/m$ it follows $ds/dP_{T} = -\rho^{-2} d\rho/dT_{P}$. 
With the thermal expansion coefficient $\alpha_T = -\rho^{-1} (d\rho / dT)_P$ this yields 
$(ds/dP)_{T} = - \rho^{-1} \alpha_T$, and usually, $\alpha_T$ is positive. 

Feature three, $ds/dP_{T} > 0$ at $\sim$6---10 Mbar, is not expected since usually $\alpha_T>0$. An exception is hydrogen, where $\alpha_T$ is known to adopt negative values in the region of pressure-induced dissociation \citep{Vorberger07,Preising23}. Figure \ref{fig:methane_entropy} shows as an example isotherms from the CMS19 hydrogen EOS \cite{chabrier19}, where $\alpha_T<0$ is seen at P$\sim$ 50---80 GPa along the 1000 K-isotherm while no longer at higher $T$ of $\gtrsim 10,000\:$K.
Could the $\alpha_T<0$ region result from methane dissociation? Both in experiments \citep{Benedetti99} and simulations \citep{Hirai2009}, methane is seen to dissociate at much lower pressure of 10--100 GPa \citep{Hirai2009,Kraus17}, similar to hydrogen, than the $\alpha_T<0$ region in our methane EoS. Furthermore, methane dissociates into  various polymers of hydrocarbons, not into atoms or ions. Therefore we conclude that feature three is not due to the dissociation of methane. 

Structural analysis of DFT-MD simulations finds that methane adopts superionic phase-like behavior at pressure of Mbars (M.~Bethkenhagen, pers.comm.~2025). According to the simulations, the carbon atoms build an amorphous lattice where their diffusivities remain higher than in the superionic phases of water or ammonia.
In this scenario, the negative $\alpha_T$ values in our methane EoS could result from the 
dissociation of hydro-carbon chains into the amorphous carbon lattice and atomic hydrogen system at several Mbar. 
Finally, it is also possible that feature three is an artifact from the applied TDI on an EoS, which is not thermodynamically consistent in that region. 

Feature four is due to the physically motivated extrapolation of the EoS where at high $P>10$ Mbar
a steep decline is also known to occur for other substances such as hydrogen or water, see \autoref{fig:methane_entropy}.

Feature five occurs for isotherms 500--2000 K and vanishes for isotherms warmer than 2500 K. The dip at 10---100 GPa could be indicative of methane dissociation into polymers, which is seen in numerical simulations to begin at 1100~K and 10 GPa \citep{Hirai2009}.
The disappearance of the dip with higher temperatures $T>2500\:$K appears to correlate with the onset of phase separation into to diamond and hydrogen as seen in compression experiments of hydrocarbons using diamond anvil cells \citep{Frost2024} or fast, dynamic shock compressions experiments \citep{Kraus17}, although the latter observe diamond formation only at higher pressures of $P>100$ GPa. However, the DFT-MD simulations underlying our methane EoS data do not resolve this phase separation, rather, an amorphous, superionic-like lattice may have formed as mentioned above.

Feature six begins to occur at $\sim$ 100 bars and may mark the regions where he ideal gas approximation breaks down. By assuming an ideal gas even at a few hundreds of bars, the internal energy and hence the entropy may be under-estimated. 
On the other hand, \citet{Ranieri24} find evidence of liquid and solid phases at T < 200 K and P up to 1  kbar in their neutron scattering experiments on methane. Liquid and solid phases are characterized by a lower entropy than in the fluid phase. Thus our methane EoS may go qualitatively into the right direction, albeit for the wrong reason.
Future work is needed to improve the methane EoS in the regions of a few 100 bars and a below 500 K. 

\begin{figure}
\centering
\includegraphics[width=0.5\textwidth]{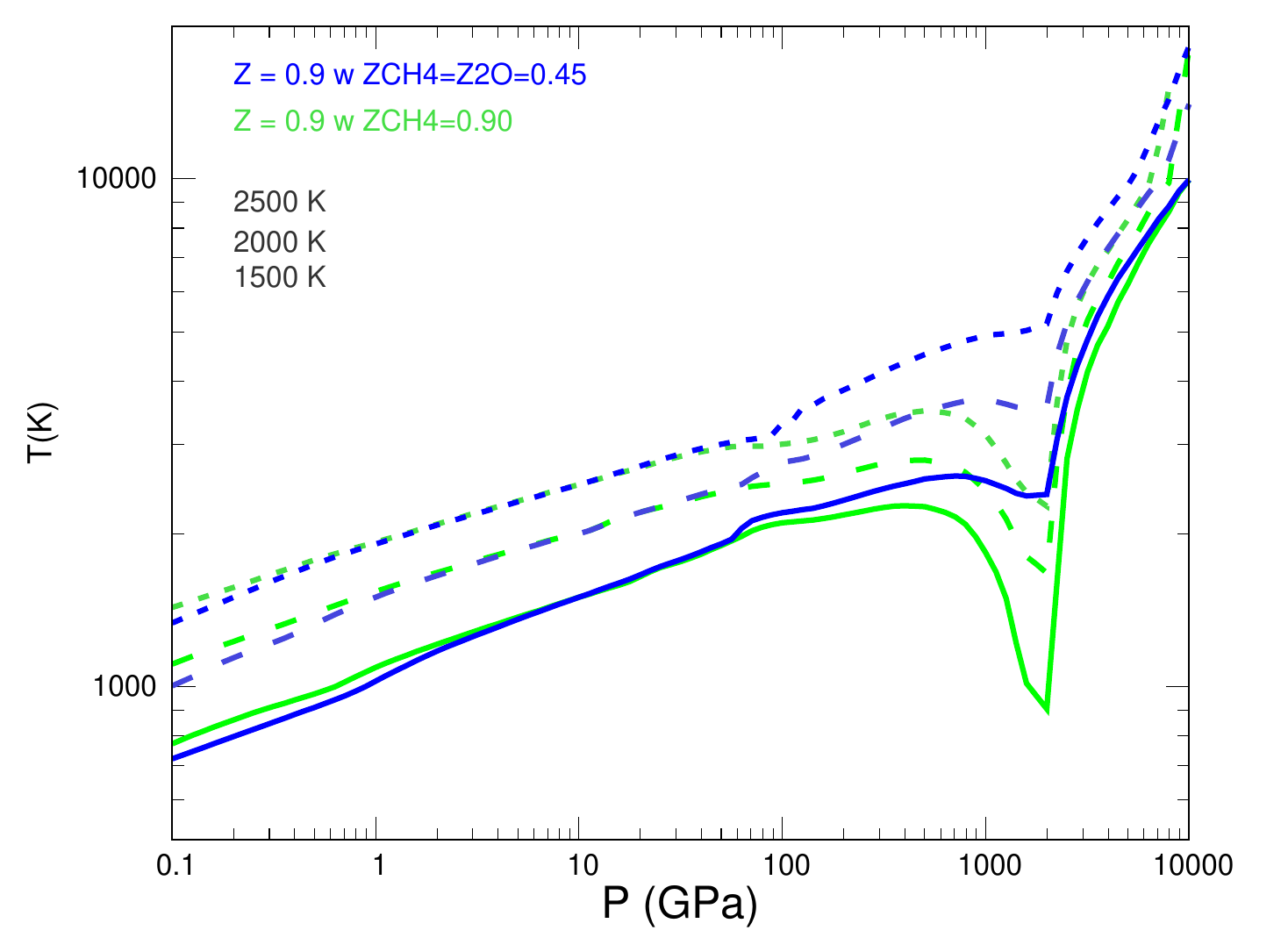}
\caption{Methane-rich adiabats characteristic of ice giant interiors. The adiabats are defined by $T$ at 10 GPa as labeled (line-style) and are computed for 90\% methane by mass (green), or 1:1 methane:water (blue), and 10\% H/He each. In the latter case, the transition to superionic phase of water appears as a shallow hill as these adiabats assume constant entropy throughout. The methane-rich adiabats exhibit the $\alpha_T<0$ behavior at high $P$.}
\label{fig:isen_methane}
\end{figure}

Figure \ref{fig:isen_methane} shows adiabats constrained by temperatures of 1500--2000 K at 10 GPa as characteristic of Uranus and Neptune for methane-rich ice giant deep interiors of 90\% methane by mass (green) or 45\% methane and 45\% water (blue). The adiabats are cold enough to adopt negative $\alpha_T$ values. While in a region $\alpha_T<0$ convection would be suppressed, the pressures of 6--8 Mbar where such a region would begin are likely too high to be of any relevance to Uranus and Neptune, where pressures at a icy mantle-rocky core transitions or $\sim$5 respectively $\sim$7 Mbar. For more massive exo-Neptunians, where interior pressures would be higher, this property of the methane EOS can be important; again we caution this feature of the methane EOS may be an artefact.

\section{Supplementary Figures}\label{apx:lumi}

\begin{figure*}[h]
\centering
\includegraphics[width=0.9\textwidth]{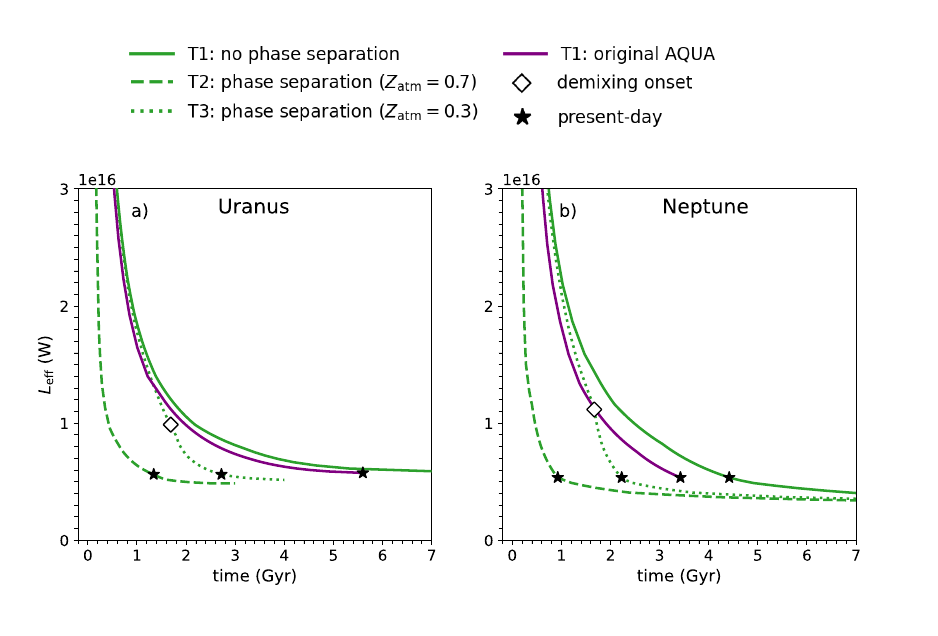}
\caption{Effective luminosity for Uranus (panel a) and Neptune (panel b) for tracks T1 (revised and original AQUA), T2, and T3. Only for T3 is the demixing onset diamond symbol seen as for T2, the onset symbol occurs outside the y-axes range.}
\label{lum-uranus-neptune}
\end{figure*}

As presented in Sections \ref{sec:results-uranus} and \ref{sec:results-neptune}, with the onset of demixing the radius increases while the effective temperatures drops. To see whether the planet becomes brighter or fainter, we show in Figure \ref{lum-uranus-neptune} the luminosity evolution of Uranus and Neptune with demixing. The luminosity decreases, hence the planets become fainter.  The reason is that the gravitational energy released from the rain-out goes into the expansion of the planet, not into an additional luminosity.

\section{Numerical accuracy}\label{apx:gravimoms} 

The automation of the interaction between the two codes TATOOINE and MOGROP is currently incomplete. At the current level of automation, using TATOOINE for the computations of the 1D-structure of a non-rotating planet, and MOGROP for the computation of the gravitational harmonics of the oblate rotating planet, introduces small inconsistencies.

One is that the contribution of rotational deformation to the gravitational and centrifugal potentials used by TATOOINE is currently neglected. To achieve good convergence between the structure and the potentials, MOGROP would normally perform seven iterations. Fortunately, this inconsistency amounts to an error in $J_2$ of only 0.04\% in
and in $J_4$ of only 0.0006\% as shown in \autoref{tab:gravimoms}.

Furthermore, some models miss the observed equatorial radius by up to 0.7\% due to the not yet fully automated interaction between the two codes. We have checked that matching the observed $\Req$ precisely rather than approximately influences the resulting cooling time by $\sim$0.1~Gyr. Fortunately, this is a small and of the same order as other uncertainties, such as neglecting rotation and the associated angular momentum conservation in the evolution runs.

\begin{table}[t]
\centering
\caption{\label{tab:gravimoms}
Convergence of the $J$'s for the Uranus-T1 model with the number of iterations between structure and gravity
computed both with MOGROP.}
\begin{tabular}{c|ccc}
  Iter.  & $J_2/10^{-6}$ & $J_4(1ß^{-6}$ & $\Req$/obs. \\\hline\hline
 1 & 3504.2285 & -35.31095 &  0.99998749 \\
 2 & 3502.3546 & -35.28558 &  0.99999164 \\
 3 & 3502.8706 & -35.29143 &  0.99999532 \\
 4 & 3502.8986 & -35.29215 &  0.99999727 \\
 5 & 3502.9226 & -35.29253 &  0.99999839 \\
 6 & 3502.9500 & -35.29292 &  0.99999906 \\
 7 & 3502.9544 & -35.29301 &  0.99999945 \\
10 & 3502.9664 & -35.29318 &  0.99999989    
\\\hline
\end{tabular}\\
\end{table}

\begin{table*}[t]
\centering
\caption{\label{tab:TMcomparison}
Inaccuracies of present-day structure models  due to the current TATOOINE-MOGROP interaction level}
\begin{tabular}{l|llllllll|l}
        & $Z_1$ & $\Zwater$ & $\Zmeth$ & $Z_2$ & $M_{core}$ & $\Req$ & $J_2$ & $J_4$ & comment \\\hline\hline
Ura-T1 & 0.141 & 0.141 & -- & 0.854 & 1.39$^{\ast}$ & 25,561 & 3509.5 & -35.9 & T1-structure by T. fed into M. \\
Ura-T1 & 0.141 & 0.141 & -- & 0.854 & 1.63$^{\ast}$ & 25,559 & 3367.3$^{\ast}$ & -33.6$^{\ast}$ & T1-abundances fed into M.\\
Ura-T1 & 0.144 & 0.141 & -- & 0.862$^{\ast}$ & 1.28$^{\ast}$ & 25,559 & 3503.0$^{\dag}$ & -35.3$^{\dag}$ & T1 by M.\\
Nep-T4  & 0.179 & 0.084 & 0.095 & 0.922 & 1.46 & 24,756 & 3529.0$^{\dag}$ & -33.9$^{\dag}$ & T4-structure by T. fed into M- \\
Nep-T4 & 0.179 & 0.084 & 0.095 & 0.912$^{\ast}$ & 1.66$^{\ast}$ & 24,766 & 3528.4$^{\dag}$ & -34.4$^{\dag}$ & T4 by M.
\\\hline
\end{tabular}\\
For the models with MOGROP (M.) only, values without symbol are taken from TATOOINE (T.), with symbol $^{\ast}$ are results, and with symbol $^{\dag}$ are fitted.
\end{table*}

Apparently, there are further differences that we do not know the origin of. 
For instance, calculating the structure with MOGROP for the same abundances $Z_1$, $Z_2$ as with TATOOINE leads to a difference in $J_2$ of 1\% and in $J_4$ of 7\% for the specific Uranus T1 model, compare the first two rows in \autoref{tab:TMcomparison}. Doing both structure and gravity consistently with MOGROP leads to differences in the resulting $Z_2$ values of 1\% while in core mass of 0.1 $\ME$ for Uranus and 0.2 $\ME$ for Neptune, 
see \autoref{tab:TMcomparison}.
These small differences are irrelevant for the evolution runs and the conclusions from this work.

\end{appendix}
\end{document}